\documentclass[12pt]{aastex}

\usepackage{emulateapj5}
\usepackage{rotating}

\shorttitle{Metallicities in Carina}
\shortauthors{Koch et al.}

\begin{document}

\title{Complexity on Small Scales: The Metallicity Distribution of the Carina Dwarf Spheroidal Galaxy 
\altaffilmark{1}}

\author{Andreas Koch\altaffilmark{2}, Eva K.~Grebel\altaffilmark{2}, Rosemary F.G.~Wyse\altaffilmark{3}, 
        Jan T.~Kleyna\altaffilmark{4},\\ 
	Mark I.~Wilkinson\altaffilmark{5}, Daniel R.~Harbeck\altaffilmark{6},  
	Gerard F.~Gilmore\altaffilmark{5}, and N.~Wyn Evans\altaffilmark{5}}
\email{koch@astro.unibas.ch}

\altaffiltext{1}{Based on observations collected at the European Southern 
Observatory at Paranal, Chile; proposal 171.B-0520(A).}
\altaffiltext{2}{Astronomical Institute of the University of Basel,
Department of Physics and Astronomy, Venusstrasse 7, CH-4102 Binningen, 
Switzerland}
\altaffiltext{3}{The John Hopkins University, 3701 San Martin Drive, 
Baltimore, MD 21218}
\altaffiltext{4}{Institute for Astronomy, University of Hawaii, 2860 
Woodlawn Drive, Honolulu, HI 96822}
\altaffiltext{5}{Institute of Astronomy, Cambridge University, Madingley 
Road, Cambridge CB3 0HA, UK}
\altaffiltext{6}{University of Wisconin, Madison, Astronomy Department, 475 North 
Charter Street, Madison, WI 53706}

\begin{abstract} The Carina dwarf spheroidal galaxy is the only galaxy of
this type that shows clearly episodic star formation separated by long
pauses.  Here we present metallicities for 437 radial velocity members of
this Galactic satellite.  The metallicities and radial velocities were
measured as part of a Large Programme with the Very Large Telescope at the
European Southern Observatory, Chile.  We obtained medium-resolution
spectroscopy with the multi-object spectrograph FLAMES.  Our target red
giants cover the entire projected surface area of Carina.  Our spectra are
centered at the near-infrared Ca\,{\sc ii} triplet, which is a
well-established metallicity indicator for old and intermediate-age red
giants.  The resulting data sample provides the largest collection of
spectroscopically derived metallicities for a Local Group dwarf spheroidal
to date.  Four of our likely radial velocity members of Carina lie outside
of this galaxy's nominal tidal radius, supporting earlier claims of the
possible existence of such stars beyond the main body of Carina.  We find a
mean metallicity of [Fe/H] $\sim -1.7$\,dex on the metallicity scale of
Carretta \& Gratton (1997) for Carina.  The formal full width at half
maximum of the metallicity distribution function is 0.92 dex, while the
full range of metallicities is found to span $\sim -3.0 <$ [Fe/H] $< 0.0$
dex.  The metallicity distribution function 
might be indicative of several subpopulations distinct in metallicity. 
There appears to be a mild radial gradient such that more
metal-rich populations are more centrally concentrated, matching a similar
trend for an increasing fraction of intermediate-age stars (Harbeck et al.\
2001).  This as well as the photometric colors of the more metal-rich red
giants suggest that Carina exhibits an age-metallicity relation.  Indeed
the age-metallicity degeneracy seems to conspire to form a narrow red giant
branch despite the considerable spread in metallicity and wide range of
ages.  The metallicity distribution function is not well-matched by a
simple closed-box model of chemical evolution.  Qualitatively better
matches are obtained by chemical models that take into account also infall
and outflows.  A G-dwarf problem remains for all of these models.
\end{abstract}


\section{Introduction}

Dwarf galaxies come in many flavors ranging from gas-rich to gas-poor, from
irregular in shape to ellipsoidal, from actively star-forming to quiescent
(Grebel 2001).  The least massive, least luminous galaxies known are the
dwarf spheroidal (dSph) galaxies.  They are characterized by absolute
V-band luminosities M$_V \ga -14$ mag, surface brightnesses of $\mu_V \ga
22$ mag arcsec$^{-2}$, H\,{\sc i} masses of $M_{\rm HI} \la 10^5
M_{\odot}$, estimated total masses $M_{tot}$ of a few times $10^7 M_{\odot}$, have
shallow projected radial light profiles and tend not to be rotationally
supported (see Grebel, Gallagher, \& Harbeck 2003 and references therein).
Because of their high velocity dispersion, their velocity dispersion
profiles, their morphology and their lack of depth extent many dSphs are
believed to be dominated by dark matter (e.g., Mateo 1997; Odenkirchen
et al.\ 2001; Klessen, Grebel, \& Harbeck 2003; Wilkinson et al.\ 2004).
In galaxy groups dSphs are usually found within $\sim 300$ kpc around more
massive galaxies (Fig.\ 1 in Grebel 2005).  The gas deficiency of dSphs
remains an unsolved puzzle -- dSphs typically contain even less gas than
expected from red giant mass loss over time scales of several Gyr. 
Searches for neutral and ionized gas usually only lead to low upper limits
on the gas content (e.g., Gallagher et al.\ 2003 and references therein).

DSphs are typically dominated by either very old populations ($> 10$ Gyr)
or by intermediate-age populations ($1 - 10$ Gyr).  Very few show more
recent star formation.  All dSphs (and all other dwarfs) studied in
sufficient detail so far have been shown to contain ancient populations
that are indistinguishable in age with the oldest age-datable populations
in the Milky Way (Grebel 2000, Grebel \& Gallagher 2004).  The detailed
star formation histories and the metal enrichment of dwarf galaxies
including dSphs vary widely; no two dwarfs are alike (e.g., Grebel 1997, Mateo 1998).
DSphs usually show continuous star formation with some amplitude variations
and declining rates at more recent times.  A number of dSphs are entirely
dominated by old populations, for instance, Draco and Ursa Minor.
 Younger and/or more metal-rich
populations in dSphs are more centrally concentrated, indicating extended
star formation episodes in the centers of their shallow potential wells
(Harbeck et al.\ 2001).  

The Galactic dSph Carina stands out among the dSphs in the Local Group
because of its unusual, episodic star formation history.  In no other dSph
clear evidence for well-separated episodes of star formation has been
found.  Carina was discovered in 1977 (Cannon, Hawarden, \& Tritton 1977)
on an ESO/SRC(J) Southern Sky Survey plate.  Carina is located at a
heliocentric distance of $94\pm5$ kpc and belongs to the fainter dSphs
($M_V = -9.4$, $\mu_V = 25.5\pm0.4$; see Grebel 2000 for references).
First indications that Carina is not a purely old, globular-cluster-like
system came from the discovery of carbon stars (Cannon, Niss, \&
Norgaard-Nielsen 1981; Mould et al.\ 1982).  Subsequent deep photometry
revealed a younger main sequence (Mould \& Aaronson 1983; Mighell 1990a,b)
and showed that the bulk of the stars in Carina is of intermediate age
(i.e., younger than 10 Gyr).  However, an old population (traced by a
horizontal branch and RR Lyrae stars, e.g., Saha, Monet, \& Seitzer 1986)
is present as well.  Mighell's (1990a,b) data show two distinct
main-sequence turnoffs.  Smecker-Hane et al.'s (1994) color-magnitude
diagram (CMD) of Carina reveals a morphologically distinct, prominent red
clump of an intermediate-age population next to a well-defined red and blue
horizontal branch.  Mighell and Smecker-Hane et al.\ concluded that Carina
must have undergone at least two separate episodes of star formation.  

The episodic nature of the star formation in Carina was then confirmed by
deeper photometry:  Smecker-Hane et al.\ (1996) presented a color-magnitude
diagram (CMD) that reached $R \sim 25$ mag.  Three well-defined, distinct
main-sequence turn-offs (MSTOs) corresponding to ages of 2, 3--6, and
11--13 Gyr are visible in these data.  Smecker-Hane et al.\ note that each
of the three MSTOs connects to the same narrow red giant branch and
conclude that ``regardless of age, Carina stars are metal-poor with [Fe/H]
$\sim -1.86$ dex and a spread in metal abundance of $\sigma_{\rm [Fe/H]}
\le 0.2$ dex.''  Mighell (1997), Hurley-Keller, Mateo, \& Nemec (1998),
Monelli et al.\ (2003), and Rizzi et al.\ (2003) arrive at similar
conclusions based on additional deep CMDs\footnote{We will not discuss here
star formation histories of Carina derived from HST photometry since the
field coverage and the number of stars are too small to do so reliably.  
}
Hurley-Keller et al.\ (1998)
argue that only 10--20\% of Carina's stars formed very early on, while the
majority formed approximately 7 Gyr ago and $\sim$ 30\% about 3 Gyr ago.
(Note that differences in the derived ages or times of the extended
episodes of star formation are mainly caused by the use of different
isochrone models).  Monelli et al.\ (2003) suggest that the blue plume
stars of Carina may be as young as 1 Gyr, and that the anomalous Cepheids
of Carina may belong to an even younger population with an age of $\sim
0.6$ Gyr (see also Mateo, Hurley-Keller, \& Nemec 1998; Poretti 1999; and
McNamara 2000 for the identification and the discussion of these stars).
The younger populations of Carina are more centrally concentrated than the
extended distribution of the old stars (Harbeck et al.\ 2001; Monelli et
al.\ 2003), similar to the trend seen in other dSphs.

Smecker-Hane et al.\ (1999) obtained Ca\,{\sc ii} triplet (CaT)
metallicities for 52 red giants in Carina.  These authors argue that the
narrow red giant branch of Carina results from an age-metallicity
conspiracy in the sense that more metal-rich, but younger stars come to lie
at the same location in the color-magnitude plane as older, metal-poor
stars, and that photometry alone underestimates the true metallicity spread
in Carina.  Based on a photometric study, Rizzi et al.\ (2003) suggest that
the narrow red giant branch is a consequence of the contribution of the
dominant intermediate-age star-formation episode, while the contribution of
the ancient episode is almost negligible.

All these findings underline the highly complex star formation history of
Carina.  There is not yet a satisfactory explanation why Carina would have
experienced episodic star formation with extended quiescent episodes in
between, and why its evolution was so different from that of other dSphs.
As pointed out by Smecker-Hane et al.\ (1996), episodic accretion of fresh
gas would be difficult given the inferred mass of Carina's dark matter
halo, and cooling and sinking back of previously ejected gas should have
led to higher enrichment.  

Our current knowledge of the detailed evolutionary history of nearby dwarf
galaxies is mainly based on photometry supplemented by rather sparse
spectroscopic information (e.g., Grebel 1999).  However, this situation is
now changing as a growing body of spectroscopic information is becoming
available.  Spectroscopy plays a particularly important role in permitting
us to break the age-metallicity degeneracy that plagues purely photometric
color-magnitude diagram analyses.  When the metallicity of individual stars
can be measured independently via spectroscopy, this information can be
added to constrain photometric determinations of the star formation
history.  For gas-deficient galaxies like dSphs, our primary source of
metallicity information are their red giants, which are now easily
accessible for ground-based 8 to 10m-class telescopes.  Ultimately, we may
be able to derive detailed age-metallicity relations.  Spectroscopic
information for a sizable number of stars is gradually becoming available
for an increasing number of dwarf galaxies (e.g., C\^ot\'e, Oke, \& Cohen
1999; Smecker-Hane et al.\ 1999; Guhathakurta, Reitzel, \& Grebel 2000;
Tolstoy et al.\ 2001; Pont et al.\ 2004).  In the current study we add
spectroscopic metallicities for several hundred red giants in the enigmatic
dSph galaxy Carina.  -- Individual element abundance ratios in stars allow
one to impose strong constraints on the modes of star formation in
galaxies.  The present study only deals with the measurement of overall
stellar metallicities.  The discussion and analysis of elemental abundances
will be left for a future paper (Koch et al., in prep.). 

Here we present a comprehensive spectroscopic study of Carina in order to
analyze the overall metal content of several hundred red giants
distributed throughout this dSph galaxy.  In the current paper we analyze
the metallicities of these stars in order to investigate several of the
questions raised in the previous paragraphs.  In particular, we wish to
measure the true, spectroscopic mean metallicity and metallicity spread of
Carina, derive its metallicity distribution function, investigate whether
age and metallicity do indeed conspire, search for possible spatial
gradients, and explore its evolutionary history taking into account its
chemical enrichment.  This paper is organized as follows: Sect.\ 2
introduces our data and describes the reduction procedures. In Sect.\ 3 we
present the steps taken to calibrate our measured CaT equivalent widths and
put them on a reference metallicity scale. The resulting metallicity
distribution function (MDF) is presented in Sect.\ 4.  In this Section
we also discuss the implications for a possible age-metallicity relation in
Carina.  In Sect.\  5, we
investigate the existence of radial gradients in the MDFs.  Sect.\ 6 then
discusses the results in the context of simple models of chemical
evolution.  In Section 7, we summarize our findings.

\section{Observations and reduction}

In the course of the ESO Large Programme 171.B-0520(A) (PI: Gilmore), which
is dedicated to the analysis of kinematic and chemical characteristics of
Galactic dSphs, we observed five fields in Carina.  These fields were
selected such that we cover most of the area of Carina, but also go a bit 
beyond its nominal tidal radius.  In these fields we obtained spectra of
red giants with the goal of determining their metallicities through the
well-established CaT method and of measuring their radial velocities to
constrain the dark matter profile of Carina.   In addition four globular
clusters (NGC\,3201, 4147, 4590 and 5904) were observed in order to permit
us to place our CaT measurements on a scale of known reference metallicities 
(Rutledge et al.\ 1997a).

\subsection{Target selection}

Targets in Carina were chosen from photometry and astrometry obtained by
the ESO Imaging Survey (EIS; see Nonino et al.\ 1999 for details) during
the Pre-FLAMES era using the Wide Field Imager (WFI) at the 2.2\,m
telescope at La Silla, Chile.  In the framework of EIS a number of
different targets and fields was observed, including a region of four WFI
fields, comprising approximately 1\,deg$^2$, located at the center of the
Carina dSph.  The reduced and transformed photometry is available from the
EIS web pages at \url{http://www.eso.org/science/eis/}.  For our
spectroscopic study, we need to pick sufficiently bright objects.
Moreover, the CaT method is only calibrated for intermediate-age and old
red giants (Cole et al.\ 2004), so we chose luminous red giant branch (RGB)
stars for the present work.  

We selected our targets to cover magnitudes ranging from the tip of the RGB
down to 3\,mag below the RGB tip, thus down to 20.3\,mag in V-band apparent
magnitude.  This corresponds to an absolute magnitude M$_V\approx$\,0.3 at
an adopted apparent V-band distance modulus of 20.05 and a reddening of
0.05 (Mighell 1997).  The actual values of the distance and reddening are
not critical for the analysis in this paper, since we utilize apparent
magnitudes relative to the horizontal branch (HB) locus.  This differential
approach also reduces the problems inherent in the EIS photometry, which go
back to intrinsic problems of WFI photometry (see, e.g., Koch et al.\
2004a,b).  Our chosen RGB magnitude range is such that even for the
faintest stars one is generally able to achieve high signal-to-noise ratios
(S/N $\sim 25$) within a reasonable integration time and under good
observing conditions (see below).  Furthermore, we selected stars across
the full width of the RGB (approximately 0.2\,mag in \bv) in order to
ensure that we were unbiased with respect to metallicity or age, and
include also potential extremely metal-poor and metal-rich giants (see the
CMD in Fig.~1).  With these constraints and with the limitations in
possible positions on the fiber plate in order to avoid fiber crossings, a
total number of 1257 stars was targeted.  However, our photometric selection is
inevitably contaminated by Galactic foreground stars.

For our calibration clusters, we used the standard fields of Stetson
(2002).  More than 80 red giant candidates altogether were picked from
Stetson's B- and V-band photometry.  These clusters range in metallicity 
from approximately $-1.1$ dex to $\sim -2$ dex in [Fe/H] (see Table~1 for
details). 

The available photometry of Carina targets is, in contrast to the Stetson
list, in the Johnson-Cousins system provided by the EIS.  Because of
differences in the filter curves, this system differs from the standard
Johnson-Cousin system as defined by Landolt's (1992) and Stetson's (2000)
{\em UBVRI} standard stars.  Conveniently, the observed EIS fields overlap
with a standard star field in Carina (Stetson 2000) allowing a direct
comparison between the EIS-photometry and that provided by Stetson. We
obtained a linear transformation using the 258 stars in common, and used
this to place both the calibration clusters and the science targets on
Stetson's homogeneous photometric standard system, avoiding any offsets
between the different sets of photometry. 

\subsection{Data acquisition}

Our observations were performed in visitor mode with the FLAMES (Fibre
Large Array Multi Element Spectrograph) facility at the Very Large
Telescope UT2 of the European Southern Observatory on Cerro Paranal in
Chile.   We observed during 22 nights spread over two semesters in 2003 and
2004 (as detailed in Table~2).  We used the GIRAFFE spectrograph with both
MEDUSA\footnote{For details on the individual components of the FLAMES
spectrograph see \url{http://www.eso.org/instruments/FLAMES/}} fiber slits
in ``low-resolution'' mode.  We observed with grating L8 centered at the
near-infrared CaT.  This set-up provides a nominal resolving power of
R=6500 at the center of the spectra and covers the wavelength range 
8206--9400\,\AA.  Each pointing is designed to handle 132 fibers,
distributed across a field of view of 25$\arcmin$ diameter.  We dedicated
about 20 fibers per configuration to blank sky for sky subtraction
purposes.

We observed five different fields in and around the center of Carina
(Table~3), which we observed with several configurations of the fiber
positioning plates in order to maximize the number of targeted stars.  The
location of our fields is displayed in Fig.~2.  The total required exposure
time per configuration in Carina was six hours in order to reach a nominal
S/N of at least 20.  This S/N is needed to obtain highly accurate
equivalent width (EW) measurements at our spectral resolution.  The
observations were split into single exposures of approximately one hour
each to facilitate cosmic ray removal. 

Most of the nights were clear and good sky conditions resulted in a typical
seeing of 0$\farcs$5-0$\farcs$7, although the first run in 2003 February
was hampered by non-photometric conditions and seeing reached values as bad
as 2$\arcsec$.  Spectra with low S/N ratios were excluded from the chemical
abundance analysis, where the lower limit for marginally measuring EWs is
at $\approx 10$, but were still useful for the kinematic analysis
(Wilkinson et al., in prep.).

\subsection{Data reduction}

In the course of ESO's standard calibration exposures, bias frames, fiber
flatfield exposures and Th-Ar spectra from a calibration lamp were taken
automatically during daytime.  We reduced our data using version 1.09 of
the FLAMES data-reduction system, {\it girbldrs}, and the respective
pipeline {\it girbldrs-pipe-1.05} (Blecha et al.\ 2000).  The basic steps
comprise bias correction as well as fiber localization and successive
adjustment via the flatfield exposures.  

Our spectra were extracted by summing up the pixel values along a virtual
slit with a width of one pixel.  The final re-binning to linear wavelength
space was obtained with the master wavelength solution from the calibration
Th-Ar spectra, which were reduced and extracted analogously to the science
spectra.  All spectra were flatfielded using the fiber flatfield exposures.
The dedicated skyfibers were employed to calculate an average sky spectrum.
This sky spectrum was then subtracted from the science spectra using the
{\sc iraf}-task {\it skytweak}\footnote{{\sc iraf} is distributed by the
National Optical Astronomical Observatory.}. The resulting accuracy (the
dispersion, in a 68\% confidence interval, of the medians of the sky
subtracted spectra in any exposure divided by the median sky) was typically
better than 2\%.

Finally, the dispersion-corrected, sky-subtracted spectra were co-added to
enhance the S/N using {\it scombine}, where the frames were weighted by
their individual S/N.  The resulting frames were then continuum-normalized
or rectified by a polynomial fit to the regions excluding the Ca and other
stronger absorption lines.  Thus the median S/N ratio obtained in this way
lies at 33 pixel$^{-1}$,  reaching $\approx\,$200 pixel$^{-1}$ for the
brightest stars.  Sample spectra of stars representative of our sample are
shown in Fig.~3.

\subsection{Membership: Radial velocities}

In order to assess the membership of each target star, we made use of the
individual radial velocities.  The determination of accurate velocities by
means of template spectra will be described in a forthcoming
paper that will concentrate on the kinematic aspects of Carina (Wilkinson
et al., in prep.).  For the moment it should suffice to note that the
Carina dSph galaxy clearly stands out against the Galactic foreground
contamination, at a systemic heliocentric radial velocity of $\sim 220$
km\,s$^{-1}$.  Membership probabilities were derived by performing an
error-weighted maximum likelihood fit of a Gaussian velocity distribution
between 150\,km\,s$^{-1}$ and 300\,km\,s$^{-1}$. Rejecting
3\,$\sigma$-outliers, the fit was iterated until convergence was reached,
resulting in a mean heliocentric radial velocity of 223.9\,km\,s$^{-1}$ and
a dispersion of 7.5\,km\,s$^{-1}$ (see Fig.~4). Taking into account our
measurement uncertainties, this agrees very well with previous results.
For example, Mateo (1998) quotes a systemic velocity of
224$\pm$3\,km\,s$^{-1}$ and a central velocity dispersion of
6.8$\pm$1.6\,km\,s$^{-1}$ (based on 17 centrally located stars).  Majewski
et al.~(2005) find a mean velocity of 222.8\,km\,s$^{-1}$ from 61 stars
that extend out to 1.5 nominal tidal radii.

Depending on the chosen probability selection\footnote{In the present
representation the 2$\sigma$, 2.5$\sigma$ and 3\,$\sigma$ limits
correspond to membership probabilities of $p>$0.0072, 0.0023 and
0.0006.}  to define a member, the total number of stars belonging to
Carina varies from 471 (2\,$\sigma$), 501 (2.5\,$\sigma$) to 510
(3\,$\sigma$). The number of high-probability member stars with
spectra sufficiently high to be useful for abundance determinations,
as opposed to radial velocity determinations, is somewhat lower,
namely 409 (2\,$\sigma$), 433 (2.5\,$\sigma$) and 437
(3\,$\sigma$).

Membership of red giants in the four calibration clusters was also
determined via radial velocities.  Only five stars (N4147-S370 and
N4147-D57 in NGC\,4147; M68-S155 and M68-S164 in NGC\,4590; and M5-1046 in
the cluster NGC\,5904) were rejected due to their strongly discrepant
($>8\sigma$) velocities.

\section{Calibration of the metallicity scale}

Determinations of accurate spectroscopic abundances require a well
calibrated and widely applicable reference scale.  The infrared lines of
the singly ionized calcium ion at $\lambda\lambda$\,8498, 8542,
8662\,\AA\,\,have become one of the spectral features of choice.  These
triplet lines are relatively easy to be distinguished, since they provide
the strongest absorption lines in the near-infrared regime of RGB spectra.
They also lie at a wavelength range where many CCDs have fairly high
sensitivity, an advantage when trying to measure faint sources within
reasonable exposure times.  However, the CaT lines are generally
considerably affected by strong night-sky emission lines, hampering
accurate measurements for abundances in faint stars.  

The use of the CaT for measuring metallicities of red giants in old,
metal-poor populations was first established for simple stellar populations
such as globular clusters (Armandroff \& Zinn 1988; Armandroff \& Da Costa
1991; Rutledge et al.\ 1997a).  It was then also employed for mixed
populations beyond the Milky Way halo, such as in dSphs (e.g., Suntzeff et
al.\ 1993, C\^ot\'e et al.\ 1999; Smecker-Hane et al.\ 1999; Guhathakurta
et al.\ 1999; Tolstoy et al.\ 2001; Pont et al.\ 2004) and in irregular
galaxies (e.g., Cole et al.\ 2005).  Based on the comprehensive catalog
of CaT data from Rutledge et al.\ (1997a), Rutledge, Hesser \& Stetson
(1997b) established a seminal calibration of Galactic globular cluster
metallicities to the scales of Zinn \& West (1984, hereinafter ZW) and
Carretta \& Gratton (1997, hereafter CG).  Nevertheless, this empirical
linear calibration is strictly valid only for old and metal poor red giants
with [Fe/H] below $-$0.5\,dex and ages no younger than 10\,Gyr, since this
is the parameter range covered by their calibrating globular clusters.

Smecker-Hane et al. (1999) carried out CaT spectroscopy of 52 red giants in
Carina and found that most of the Carina RGB stars belong to a metal poor
population around $-$2\,dex.  Still, the occurrence of a wide range of
ages, particularly the dominant intermediate-age populations (Hurley-Keller
et al.\ 1998; Rizzi et al.\ 2003), hence raises the question of an
unbiased applicability of the CaT-technique to our Carina sample.
However, recently the CaT calibration was extended to ages as young as 2.5
Gyr:  Based on spectroscopy of red giants in Galactic globular and open
clusters Cole et al.\ (2004) have shown that the CaT method has very little
age sensitivity and can be reliably applied to composite populations with
a range of ages.   Since the dominant populations in the Carina dSph lie in
the range of validity of the CaT method, we can trust that our measurements
will yield consistent results for this galaxy.

\subsection{Equivalent widths and reduced width}

Equivalent widths of the three CaT lines were measured by means of the
software kindly provided by A.\ Cole, which is a modified version of G.\ Da
Costa's {\sc ewprog}-program (see Cole et al.\ 2004). We followed Suntzeff
et al.\ (1993) in using line and continuum bandpasses as defined in
Armandroff \& Zinn (1988). These were shifted to take account of our
derived radial velocities.  As was proposed by Cole et al.\ (2004), each
CaT line is best fit by the sum of a Gaussian and a Lorentz profile with a
common nominal line center, which yields a progressively better fit to the
line wings. Theoretical models show that the wings are most sensitive to
changes in stellar parameters (e.g., Smith \& Drake 1990).  The final width
was then derived by summing up the flux in the fit profile across the
bandpass and errors were calculated automatically from the residuals of the
fit.

In order to perform a self-consistent analysis of our calibration clusters
and of the Carina data, we have to re-derive the relation between line
strength and metallicities. We chose to exclude the weakest of the three
lines at 8498\,\AA\ in order to minimize the errors in the final
metallicity measurements. Thus we define the Ca linestrength as

\begin{displaymath}
\Sigma W = EW_{8542} + EW_{8662}. 
\end{displaymath}

The overall width $\Sigma W$ of the CaT lines is not only a strong function
of metallicity, but is also affected by stellar effective temperature and
surface gravity. Isolating the strong sensitivity of the linestrength to
metallicity is commonly achieved by defining a {\it reduced} equivalent
width, which effectively removes the signature of surface gravity and
effective temperature from the CaT strength (Rutledge et al.\ 1997b and
references therein).  Basically, correcting the linestrength for a star's
V-magnitude overcomes the sensitivity to gravity, since a given star on
the RGB can have a range of radii depending on its evolutionary phase,
although its mass remains nearly constant.  The increase in radius toward
the tip of the RGB results in a progressively lower surface gravity (and
effective temperature), where the relation between gravity and V magnitude
is close to linear (e.g., Cole et al.\ 2000). The additional introduction
of the apparent magnitude of the horizontal branch, V$_{HB}$, and the
subsequent measurement of the difference between the apparent V-band
magnitude of a given red giant and of V$_{HB}$ removes any remaining
dependence on systemic distance and reddening.  

As a comparison with the maps of Schlegel, Finkbeiner \& Davis (1998)
reveals, the Galactic foreground extinction toward the region of Carina is
rather constant (with an r.m.s.\ scatter around the mean E(B$-$V) of less
than 0.01\,mag across the projected area of the galaxy).  Thus differential
reddening is unlikely to significantly falsify the use of a single HB
magnitude throughout this dSph.  In a gas- and dust-deficient dSphs like
Carina the internal reddening may be neglected.  Furthermore, since the
color range across the RGB is small and since we are using Johnson-Cousins
colors, any extinction-induced magnitude differences for stars of different
surface temperature may be entirely ignored (Grebel \& Roberts 1995).
Hence the reduced width, defined as
 
\begin{displaymath}
W^{\prime} \,=\, \Sigma W \,+\, \beta\,(V\,-\,V_{HB}),
\end{displaymath}

\noindent 
is basically the CaT-linestrength of an RGB star with the luminosity that
it would have if it were located on the HB. This parameter is a very
convenient metallicity indicator (Rutledge et al.\ 1997b), since in this
form it is {\em purely} a function of metallicity.
 
Fig.~5 shows a plot of the measured linestrengths for each of the
calibration clusters versus the respective magnitudes above their HBs.  We
adopted the HB magnitudes of our calibrators from Ferraro et al.\ (1999) in
order to place all measurements on a consistent scale.  Likewise, we
defined the HB level of Carina via the magnitude of the lower envelope of
the observed HB distribution in the region with 0.2\,$<\bv<$\,0.6.  Hence
we consistently use the determined V$_{HB}$ magnitude of 20.88\,mag.  This
is the HB of the {\it old\/} population in the Carina dSph; the
core-Helium-burning stars of the (dominant) younger/intermediate-age
population form the prominent red clump at $V \sim 20.5$ (see Monelli et
al.~2003; Smecker-Hane et al.~1994).  As was demonstrated by Cole et al.\
(2004) using the HB location of the old HB is perfectly acceptable also for
intermediate-age populations.  We discuss the (small) expected amplitude of
the resulting systematic effect in the metallicities of the
intermediate-age stars below.
  
A linear, error-weighted least-squares fit to the line measurements of all
remaining 72 stars in the four globular clusters leads to the
parameterization of the reduced width, W$^{\prime}$, as 

\begin{equation}
W^{\prime} \,=\, \Sigma W \,+\, (0.55\,\pm\,0.02)\,(V\,-\,V_{HB}),
\end{equation}

\noindent
with an r.m.s.\ scatter around this averaged slope of 0.18\,\AA.  In this
equation, the given slope is the error-weighted mean of the individual
slopes of each of the four clusters.  The value for the slope derived by us
is shallower than the one obtained in the original calibrations by Rutledge 
et al.\ (1997a), who measured it to be (0.64$\,\pm\,$0.02) \,\AA\,mag$^{-1}$
based on the CaT line strengths in 52 Galactic globular clusters.  The more
recent analysis of Cole et al.~(2004), who included metal-rich open
clusters, obtained an even steeper slope of $0.73\pm 0.04$. They suggest
that the difference is due to an increase of slope with the mean
metallicity of the calibrating clusters. Indeed, the mean metallicity of
our four calibrating clusters, at $-1.47$~dex on the CG scale ($-1.70$~dex
on the ZW scale), is $\sim 0.2-0.4$~dex more metal-poor than the entire
sample used by Rutledge et al.~(1997a,b) (mean metallicities of $-1.25$~dex
on the CG scale, $-1.33$~dex on the ZW scale, resp.), giving some support
to the interpretation of Cole et al.~(2004). Our calibrating clusters were
selected to cover the expected metallicity range of the Carina dSph.
Further, our analysis is based on only the two stronger CaT lines and uses
a different line profile to that used by Rutledge et al.\ (1997a). 
In this paper, we shall consistently use our own calibration.

\subsection{The metallicity scale}

One of our observed globular clusters, NGC\,4147,  was not included in the
common scale of Carretta \& Gratton (1997). In order to exploit the maximum
number of calibrators at hand, we nevertheless retained this cluster
in our analysis by adopting the reference metallicities by Rutledge et al.\
(1997b), who also measured this cluster.  These authors calibrated their
sample of globular clusters to both the ZW and CG scales, and we also apply
calibrations to both scales for the sake of consistency. 
Another calibration scale is provided by Kraft \& Ivans (2003), who 
derive globular cluster metallicities based on the equivalent widths of 
Fe\,{\sc ii} lines from high-resolution spectroscopy. Their scale yields 
metallicities that are 
about 0.2 dex more metal poor than those on the CG scale. 
However, we will
present our results later on in terms of the calibration of Rutledge et
al.\ (1997b) on the scale of CG unless stated otherwise, since this is the
most common representation nowadays.

As Fig.~6 implies our observations yield well-defined linear relations
between the reduced width W$^{\prime}$ and [Fe/H] on both scales.  We can
thus infer the metallicities of our target stars using the following
equations, which were obtained by an error-weighted least-squares fit:

\begin{eqnarray}
\mathrm{[Fe/H]}_{R97,CG} \,&=&\, (-2.77 \,\pm\, 0.06)\,+\,(0.38\,\pm\,0.02)\,W^{\prime}\\
\mathrm{[Fe/H]}_{R97,ZW} \,&=&\, (-2.74 \,\pm\, 0.06)\,+\,(0.30\,\pm\,0.02)\,W^{\prime}
\end{eqnarray}

The bottom panel of Fig.~6 shows that there is only little scatter around
the best-fit line.  The r.m.s.\ scatter was calculated to be 0.02\,dex for
both metallicity scales. The uncertainties of our metallicities were estimated by
propagation using the measurement errors in the EWs and
uncertainties in the calibration (eqs.\ 1--3). The mean measured
uncertainty in our metallicities amounts to 0.17\,dex.  

Red giants that were observed in Carina are shown
in the V,$\Sigma$W-plane of Fig.~7, together with some isometallicity lines
to guide the eye.  The wide range of values in the diagram already
indicates a large spread in metallicities.

\subsection{The possible impact of Ca abundance variations}

The ``metallicity'' of a galaxy is customarily represented by its iron
abundance [Fe/H] (or [Me/H]).  Strictly speaking, Ca does not necessarily
trace Fe as such, and Ca and Fe are formed in different nucleosynthesis
processes.  A potential inconsistency in applying the CaT technique
to derive metallicities in a dSph in terms of [Fe/H] is the likely
variation of the [Ca/Fe] ratio, both within the composite population of the
dSph, and in between the calibrating clusters and the dSph stars. When
converting from Ca to Fe abundances, one has to consider the different
dependencies of [Fe/H] and [Ca/Fe] on the detailed star formation histories
of the galaxy. It is far from straightforward to determine an unambiguous
metallicity scale from single element measurements, particularly in a
galaxy with a strongly episodic star formation history like Carina (see
Cole et al.\ 2000; Tolstoy et al.\ 2001 for a detailed discussion and
Gilmore \& Wyse 1991 for models).  

The measured [Ca/Fe] ratios for stars in globular clusters tend to be
higher than the solar value (see the compilation in Carney 1996). Indeed,
the values for three of our calibrating clusters (each based on only a
handful of stars; Carney 1996) are (see also Table~1): 
${\rm [Ca/Fe]} = +0.11 \pm 0.04$
(NGC~3201); ${\rm [Ca/Fe]} = +0.32$ (NGC~4590; only two stars); ${\rm
[Ca/Fe]} = +0.21 \pm 0.02$ (NGC~5904).  The few published elemental
abundances for stars in the Carina dSph show some stars with similar
values, but a tendency toward the lower values as expected for an
extended star-formation history and incorporation of Fe from Type Ia
supernovae.  Shetrone et al.~(2003) analyzed five stars that are known to
be members from the radial velocity measurements of Mateo et al.~(1993).
They obtained values for [Ca/Fe] of: +0.14, $-0.10$, +0.20, +0.12 and
$-0.02$, with a typical uncertainty of 0.05 (see Table~4). 
 The values of [Fe/H] that
were based on Fe lines in Shetrone et al.'s (2003) high-resolution spectra
for these bright 
stars are: $-1.60,\, -1.64,\, -1.60,\, -1.40$ and $-1.94$
respectively.  

We note that uncertainties resulting from the unknown variations in the
values of [Ca/Fe] for our target stars are likely to be up to 0.2~dex,
based on these measured values.  
Such effects may in the future be overcome by calibrating W$^{\prime}$ 
directly onto [Ca/H] (Bosler et al.\ 2002), making calibrations 
independent of star formation history, unlike the present technique.   
Our own high-resolution UVES spectra for
approximately 30 stars will provide more elemental abundance information,
which can be used to obtain an improved statistical estimate of the
metallicity uncertainty (Koch et al., in prep.).

The high-resolution measurements from Shetrone et al.\ (2003) quoted above
in fact also provide an external check of our metallicity estimates. 
Comparing their results with our data permits us to assess the 
internal precision of our measurements. 
Table~4 gives a comparison of [Fe/H] values from our analysis with the
respective high-resolution data. The 
mean 
offset between both analyses
is 0.09\,dex (our CG scale) and 0.13\,dex (ZW). 
After adjusting the individual deviations by these mean offsets, the 
absolute average difference amounts to 0.18 (0.25) dex.  This offset is 
comparable to the order of uncertainty introduced by the [Ca/Fe] variations 
with respect to Galactic globular clusters as mentioned above, as well as 
our random measurement errors.  Since we here juxtapose our Ca-based 
metallicities with the true iron abundances from Shetrone et al.\ (2003), 
the above quoted average difference well reflects the calibration uncertainty. 

\subsection{Additional potential sources of uncertainty}

Furthermore, while ideally one would like to associate each RGB star with
the appropriate HB level corresponding to a population of the age and
metallicity of that RGB star, in practice we do not have sufficient
information. Our adoption of a single HB for all the stars has the
advantage of providing us with a well-defined magnitude and allows us to
estimate the associated uncertainties.  Since we adopted the HB appropriate
for the oldest stars, and hence the least luminous HB, there will be a
systematic effect.  The expected result from normalizing with an HB of
apparent magnitude $V_{HB,true} + \Delta V_{HB}$, with $V_{HB,true}$ the
appropriate HB level for a given star, can be seen by the appropriate
substitution in the relations for reduced width and [Fe/H] (eqs.1,3).  We
find a resulting error in [Fe/H] of $-0.55 \times 0.38 \times \Delta
V_{HB}$~dex.  As noted earlier, the core He-burning phase of the
intermediate-age population in the Carina dSph forms a prominent red
clump, some 0.4~mag brighter than the old HB used in the normalization;
thus for these stars $\Delta V_{HB} \sim 0.4$. The resulting bias in
[Fe/H] is $-0.07$~dex, i.e., we estimate that these stars may be
$-0.07$~dex too metal-poor.  

The small amplitude of this effect is confirmed by the analysis of Cole et
al.~(2004), where they experiment by assigning HB levels (absolute
magnitudes) randomly to stars in different clusters.  The variation in the
luminosity of the clusters' HB levels is 0.4~mag (see their Table~1).
Using their derived slopes for the calibration relations, we find that the
total amplitude of the change in metallicity estimates from using the
appropriate HB level is 0.1~dex.  This is just what is seen in their
Figure~6.  Basically the variation of HB level with age and metallicity
amounts to $\sim 0.1$~dex in metallicity uncertainty, and in our case for
most of the stars results in an underestimate by this amount. Of course we
cannot say for which stars we are underestimating the metallicity without
having age information. As we discuss below, in principle, with robust
accurate models of the upper RGB, one could iteratively solve for age and
metallicity.  We did not attempt this procedure since the maximum amplitude
of the effect, $\sim 0.1$~dex, is within the uncertainties of the models.

Of the target stars in the central field, 46 were observed repeatedly,
i.e., during all of the runs.  In these cases the finally quoted
metallicity is the value derived from a combined spectrum, including all of
the respective single spectra. However, in order to get an estimate of the
random errors of the measurements, we compare in Fig.~8 the metallicities
determined from spectra obtained on the individual observing runs in 2003
February and March to metallicities derived from combined spectra from all
the available observations (see also Table~5).  Some of the entries have
metallicity estimates from the two runs differing by several sigma.  It
turns out that the median difference ([Fe/H] (1st run) $-$ [Fe/H] (2nd
run)) between individual metallicity estimates amounts to 0.08\,dex 
and the 
r.m.s.\ scatter of the different estimates from both runs totals 0.57\,dex. 
Rejecting those objects deviating more than two standard deviations, 
namely two targets with particularly low S/N, 
from the sample we find a respective r.m.s. of 0.3\,dex.
This is larger than our overall random error estimate of 0.17\,dex. But 
as noted before, this is mainly due to the first run being plagued by 
poor weather conditions, which was taken into account by weighting spectra 
by their S/N ratio during the co-addition.
Consequently, the mean uncertainty of 0.24\,dex for the targets observed 
during the first run compares to 0.13\,dex for the same objects targeted 
in the second run.

\section{The metallicity distribution}

Metallicities derived for each individual star are given in Table~6.
The resultant MDFs (see Fig.~9) turn out to be almost indistinguishable 
when using 
different membership selections (i.e., cutting at different $\sigma$ levels 
in the radial velocity distributions), except for the varying total number.  
We used both a K-S test and 
the Kuiper statistics (Press et al.\ 1992) to test how different 
the MDFs are for different selection criteria.  The Kuiper statistics
is more sensitive to differences in the MDFs' tails. 
The MDFs resulting from a $1\sigma,\,2\sigma$,  and 3$\sigma$-cut 
are consistent with each other at a level of greater than 99.9\%.  
Hence we chose to maintain the limit of 3\,$\sigma$ as our
membership criterion.

\subsection{The width of the MDF}

Histograms of the resulting metallicities on each of the two  calibration
scales for the sample of the 437 Carina red giants are displayed in
Fig.~9.  These show a broad distribution, peaked at a mean metallicity of
$-1.72\pm 0.01$\,dex (CG scale) or $-1.90\pm 0.01$ (ZW scale),
respectively.  The peak of the MDF is, however, more pronounced for the case of 
the CG-scale.
 The entire distribution's {\it formal} full width at half
maximum is 0.92\,dex.  The full metallicity range is $\sim$~3.0~dex, as
can be seen in Fig.~9.  
This spread can in part be attributed to the measurement uncertainties, 
but also an
occurrence of several subpopulations with different peak metallicities 
cannot be rejected as a further source of the broadening (see below).

The formal 1\,$\sigma$-width of the distribution is 0.39\,dex, but since a
best-fit single Gaussian is simply a mathematically convenient description,
some of this width will also be due to measurement and calibration
uncertainties.  Subtracting our best estimates of these, namely 0.17\,dex 
for the measurement error and 0.02 uncertainty in the slope of the
calibration in quadrature, we estimate the width of our distribution to be 
$\sigma\,=\,$0.33\,dex. 
One should keep in mind, however, that the combined
effects of calibration uncertainties from the [Fe/H] calibration via
reference clusters through systematic effects due to varying [Ca/Fe] and
normalizing the HB level may amount up to 0.22\,dex in total. 
Accounting also for these systematic errors, 
we obtain an estimate of the width of the MDF of $\sigma\,=\,0.25\,$dex.
Furthermore, we recall that the dominant population
in Carina is of intermediate age, reducing the possibility of star-to-star
random variations of uncertainties introduced by the chosen HB level.  Thus
the width of the distribution is in fact dominated by physical variations
in metallicity. 

In the extreme tails of the MDF we find stars with metallicities
approaching $-3$ dex and near-solar metallicity, respectively.  Note that our
calibrating globular clusters only cover the initially expected range of
metallicities in Carina, i.e., metallicities from $\sim -2$ to $\sim -1$
dex (Table~1).  Hence the metallicities for the stars in the tails of the
MDF are based in extrapolation of our calibration.  Follow-up spectroscopy
of these stars would be desirable in order to check whether this
extrapolation yields reliable results and to further elucidate the chemical
properties of these stars.  If the overall [Fe/H] values for these stars
are correct, they would be as metal-poor as the most metal-poor red giants
found in other nearby dwarf galaxies (e.g., Shetrone et al.\ 2001).  At the
metal-rich end, a handful of them would be as metal-rich as the metal-rich
population in the Sgr dSph (Bonifacio et al.\ 2004).

\subsection{Comparison with previous studies}

We can now compare the results of our large sample with previous results
for large samples of red giants in Carina.  {From} a sample of 52 RGB stars
using the same method as here, Smecker-Hane et al.\ (1999) found the mean
spectroscopic metallicity of Carina to be $-1.99\,\pm\,0.08$\,dex.  These
authors have not published individual metallicities, so we cannot perform a
direct comparison between stars in common to both samples.  Rizzi et al.\
(2003) analyzed the color distribution across the RGB, including a
statistical correction for the star-formation history inferred from the
main sequence turn-off region.  They derived a photometry-based average
metallicity of $-1.91\,\pm\,0.18\,$dex.  These results on Carina are in
reasonable agreement with our data if the quoted uncertainties and the
widths of the distributions are taken into account.

Yet, metallicity estimates via  purely photometrical isochrone fitting of
RGB stars without age information lead to mean values that are typically
more metal-rich by approximately 0.2\,dex, as, e.g., in Monelli et al.\
(2003).  These authors obtain their result by a direct comparison with the
Galactic globular cluster NGC\,1904, a single-age, single-metallicity
system.  For this object, different values for [Fe/H] are found in the
literature, reaching from $-1.37$\,dex (Rutledge et al.\ 1997b on the CG
scale) to $-1.69$\,dex (loc.\ cit., ZW-scale), whereas Harris (1996) cites
an average metallicity of $-1.57$\,dex from all available sources.  Still,
one has to keep in mind that such a purely photometric fit to the RGB,
without any input from an age distribution and age-metallicity relation, is
subject to large uncertainties, as is underscored by Fig.~10.  An
additional source of uncertainty is the ability (or lack thereof) of many
isochrone models to correctly reproduce the slopes of the red giant branch
for very low or high metallicities (e.g., Fig.\ 5 in Grebel 1997 and Fig.\ 5
in Grebel 1999). 

\subsection{ The shape of the MDF}

The complex morphology of the CMD of the Carina dSph -- both the multiple
turn-off regions and the different He-burning phases -- is compelling
evidence for strongly varying star formation rates, in fact even for
distinct episodes.  The metallicity distribution of low-mass stars, as
derived here, is rather insensitive to the details of the star formation
history, particularly if elements dominated by massive stars are used
(cf.~Pagel 1997, Section 8.3.7).  However, we have calibrated our CaT
metallicities onto {\it iron\/} (assuming the same [Ca/Fe] in the calibrators
as in our stars), which in systems with extended star formation can have a
significant contribution from long-lived stars through Type Ia supernovae.
Thus one can have enrichment in Fe without accompanying star formation.
Hence with a star-formation history that consists of bursts of star
formation with a real hiatus between the bursts, the Fe distribution can
have ``gaps'' (no stars formed while Type Ia supernovae enriched the gas,
prior to the next burst; cf.~Gilmore \& Wyse 1991).  
In order to test whether any substructure reflected in the MDF may 
provide evidence of star formation events, 
we applied the Shapiro-Wilk test for normality of a distribution (e.g., 
Royston 1995) to our observed binned MDF. As a result we can reject the hypothesis of an 
underlying single Gaussian normal distribution with a significance of 96\%. 
On the other hand, the observed MDF is not expected to be strictly Gaussian, 
since star formation events do not naturally produce such normal distributions. 
In addition, we applied a KMM test for multiple populations imprinted in the 
MDF (Ashman, Bird, \& Zepf 1994) 
to our  unbinned data. Again, given our measurement uncertainties and the non-normality of 
astrophysical star formation events, one has to keep in mind that such a decomposition into multiple
single Gaussian sub-populations is purely a formal procedure at this stage
that does not necessarily have a physical meaning.
We find that the case of four
populations is significantly preferred against a
one-population model at the 98.1\% confidence level.  Such a potential
multi-population
star formation history could also easily account for the MDF's metal-rich
extension.  
Yet a proper interpretation of the observed metallicities requires detailled modeling of the 
chemical evolution to which we will return in Sect.~6 below. 

\subsection{An age-metallicity relation?}

The (lack of a) relation between the color of our targets and their
metallicity is demonstrated in Fig.~10.  This figure shows that there is no
obvious trend in the sense that predominantly metal-poor stars are not
concentrated along the blue ridgeline of the RGB, and the metal-rich stars
are not concentrated at the red ridgeline (in contrast to the LMC data of
Cole et al.\ 2005, where such a trend is clearly visible).  A
metallicity-color trend would have been expected to result from a direct
comparison with fiducials of globular clusters of known metallicities,
i.e., for populations of roughly the same age but different metallicities,
as indicated by the RGB fiducials for four globular clusters of Sarajedini
\& Layden (1997) in Fig.~10 (left panel). The ridgelines from Sarajedini \&
Layden cover a wide range of colors in the CMD, corresponding to a
metallicity range from $-2.02$ to $-0.78$\,dex  
for old and nearly coeval
populations.  On the right-hand side we show three isochrones with ages and
metallicities that roughly correspond to what may be expected from the
three main star formation episodes in Carina.  Again there is no clear
correlation with the data points.  Hence it is clear that one cannot derive
the metallicity of an individual star from its color and magnitude on the
RGB when dealing with a mixed-age population as present in Carina, nor can
a metallicity distribution directly be inferred from the color
distribution in this case.

The existence of an age-metallicity relation can be inferred from the fact
that the giants appear to be equally widely spread across the RGB
regardless of their metallicity.  At a given age, metal-poor stars are in
general bluer than their metal-rich counterparts owing to the reduced
opacity in their photospheres.  Conversely, the older stars tend to have
redder colors at a given [Fe/H].  Hence, in order to produce the
counteracting trend with both metal-rich and metal-poor stars at the same
locus in the CMD, the metal-poor stars necessarily have to be older,
whereas the more metal-rich giants have to be of younger age. 

If one had reliable models of the upper RGB in the observational plane, one
could in principle derive age estimates for our stars of known metallicity,
from finding the best-matched isochrone at that metallicity.  However, the
small offset in color with age at the relevant intermediate ages, combined
with sensitivity to the largely unknown elemental abundance mix and He
content limits the usefulness of this exercise. An age distribution and
age-metallicity relations are better derived from turn-off stars, provided
that the photometry is at a sufficient level of accuracy.  Such an analysis
with high-resolution spectra will be subject of a forthcoming paper (Koch
et al., in prep.).

\section{Radial variations}

Previous analyses based on wide-area photometric surveys and thus derived
CMDs have shown that the stellar populations in the Carina dSph feature
different spatial distributions.  Mighell (1997) proposed a strong central
concentration of intermediate-age stars from a comparison between a deep
HST CMD of the center and ground-based data that covered a wider area.
This was confirmed by Harbeck et al.~(2001) via wide-field photometry.
Harbeck et al.\ found a pronounced central concentration of the prominent
intermediate-age red clump stars in comparison to the old HB.  This was
subsequently also seen in the photometric study of Monelli et al.~(2003).
Harbeck et al.\ (2001) discarded the presence of a significant radial
gradient in Carina's old HB morphology out to its nominal tidal radius. 

Our large dataset allows us to address the issue of population gradients by
looking for radial gradients in the 
MDFs.  For this purpose we investigated the MDFs as a function
of elliptical radius $r=[x^2+y^2/(1-\varepsilon)^2]^{1/2}$, where $x$ and
$y$ denote the distances along the major and minor axis,
respectively\footnote{Here we adopted a position angle of
(65$\,\pm\,5)\degr$ and an ellipticity of $\varepsilon\,=\,0.33\,\pm\,0.05$
(Irwin \& Hatzidimitriou 1995).}.
There is no obvious strong radial gradient in the resulting scatter plot of
metallicity against elliptical radial coordinate, as shown in Fig.~11 (top
panel).  A simple linear fit yields a gradient of
($-0.0047\pm0.0008)$\,dex\,arcmin$^{-1}$.

However, as Fig.~11 (bottom left) implies, there is a clear radial tendency
in the respective MDFs to be seen when the data are split into three radial
bins.  The annuli have been chosen to cover everything within one core
radius (r$_c=8\farcm$8), then everything within r$_c$ to 2r$_c$, and from
2r$_c$ to the nominal tidal radius at 28$\farcm$8.  The ratio of the total
number of stars in the annuli is approximately 5:4:1.  Fig.~11 shows the
single MDFs, represented by normalized generalized histograms, which were
obtained from the density distributions of our [Fe/H] measurements by
convolution with a Gaussian distribution accounting for the individual
measurement errors. 

The MDF of the innermost region is already rather broad with a suggestion
of both lower and higher metallicities and 
an indication of a 
secondary population around 
approximately $-2$\,dex. However, the
MDF of the middle annulus is peaked at a metallicity that is at least
0.1\,dex lower than in the inner region and shows an increase of the 
potential 
secondary peak at $\sim -2$\,dex.  
Considering the overall uncertainties in the metallicity measurements, 
a shift by,
e.g., 0.1 dex may not appear significant.  However, if we take into account that
the uncertainties of the metallicity measurements should be the same
everywhere in Carina and that there is no obvious reason why these
uncertainties should introduce systematic offsets from one bin to another,
these differences may be meaningful. 

A K-S test revealed a probability of
98.2\% that the MDF from the innermost ring is drawn from the same parent
population as the entire sample. 
Based on the same statistical test, we cannot reject the hypothesis 
that also the middle annulus is drawn from the same population as the 
innermost annulus and the entire sample, resp. (at significances of
35\% and 50\%).

The MDF of the outermost region, on the other hand, shows several peaks,
the broadest and most distinct of which is more metal-poor by
$\sim$0.3\,dex than that the one seen in  the innermost region.  In
contrast, the location of a second peak at an [Fe/H] of $-$1.6\,dex
is consistent with the MDFs' peaks in the inner parts of the galaxy to
within the measurement uncertainties.  
Using a K-S test we find that the probability that the outer and inner MDFs are
drawn from the same distribution is less than 4\% and we can therefore 
reject this hypothesis. Moreover, the
probability of the outer region being compatible with the entire sample is
at the 6\% level. 
Hence there do seem to be indications of spatial (and
indeed radial) variations in Carina's stellar populations.

This gradient is also reflected in the cumulative metallicity distributions
shown in the bottom right panel of Fig.~11. In a complementary manner, a cumulative plot of
the giants' galactocentric radii clearly indicates a preferential central
concentration of the metal-rich component (Fig.~12) compared to the more
metal-poor population --  a K-S yields a probability that the metal-poor and
metal-rich populations take the same spatial distributions of 10\%.   

The evidence of a concentration of the metal-rich component in Carina that we found 
is consistent with earlier findings that the intermediate-age
stars favor the inner regions,
 in particular if one takes into consideration that we have likely underestimated their
metallicity  by $\sim 0.1$~dex due to the adoption of the old HB in the
reduced CaT width.  Thus, if the metal-rich component can be attributed to
the intermediate-age population and, vice versa, the metal-poor RGB stars
in Carina are to be identified as predominantly belonging to an old
population, then the radial variations and progressive shifts of the peaks
found in the spectroscopically determined MDFs underscore previous findings
of centrally concentrated starbursts based on CMD analyses (Harbeck et al.\
2001). A reason for such a concentration could be stronger cooling and
dissipation of more metal-enriched gas.   It may also indicate that
star-forming material can be more easily retained at the center of a dSph's
shallow potential well.

\section{Implications for chemical evolution} 

As the present-day observed MDF of a stellar system reflects its entire
integrated history it is relatively insensitive to the star-formation
history (SFH).  The SFH cannot be unequivocally derived unless these data
are combined with accurate ages (and preferably also individual elemental
abundances to constrain the actual modes of star formation, see, e.g.,
Shetrone, C\^ot\'e, \& Sargent 2001).  However, the gross properties of the
MDF can provide insight into certain aspects of the chemical evolution of a
system, such as the importance of gas flows. Following long-standing
practice, we gain a basic insight by consideration of variants of the
simple model of chemical evolution (e.g., Pagel 1997, Chapter 8) and refer
the detailed stochastical modeling to a future paper (Wyse et al., in
preparation).

In such basic models, the mean metallicity of long-lived stars at late
stages (close to gas exhaustion) is representative of the true
nucleosynthetic yield, $p$, if all metals are retained in the system.  If
the system is not closed and if outflows are possible, the mean metallicity
reflects the `effective yield'', a combination of true yield and effects of
loss of metals (see Hartwick 1976): outflows at a constant factor $c$ times
the star-formation rate reduce the mean metallicity below the true yield by
approximately this factor $c$, $p_{\mathrm{eff}} \sim p/c$.  Thus, assuming an
invariant stellar initial mass function (and thus invariant true yield
$p$), the low peak metallicity of our derived MDF, compared to say the
G dwarfs in the solar neighborhood, which peaks at $\sim -0.2$~dex
(e.g.,~Nordstr\"om et al.~2004 and references therein) suggests a strong
influence of outflows, presumably driven by Type II supernovae. Indeed,
this has been the basic model for the evolution of gas-poor dwarf galaxies
since Sandage (1965).  

\subsection{Outflows}

A modified simple model with flows overpredicts the number of metal-poor
stars (leading to a ``G-dwarf problem'' just as in applications of this
model to the solar neighborhood). 
Having normalized the fit of our models in Fig.~13 to the 
total number of stars observed, the discrepancies between the prediction and 
observations are marginal at $-2.5$\,dex, but become significant below $-2.7$\,dex, 
where the deviations progressively exceed the $2\,\sigma$-level.
The fit may be improved by adopting some
non-zero initial metallicity (or prompt initial enrichment (PIE), e.g.,
Tinsley 1975), understood in the present-day context as being from
pre-galactic Population III stars.  The predicted metallicity distributions
for a simple model with outflows, and with PIE, are shown compared to the
full Carina data set in Fig.~13. 
The respective best-fit parameters to our data are $p_{eff}=0.028\,Z_{\sun}$ 
for the simple model, and $p_{eff}=0.029\,Z_{\sun}$ and $Z_{init}= 0.0013\,Z_{\sun}$ 
in the case of PIE. 
One should note that these models
retain the instantaneous recycling approximation and so are not expected to
provide a perfect fit to our iron-based MDF.  Further, the radial
variations discussed in Section 5 imply that the system is not in fact
well-mixed at all times, another assumption inherent to the simple model.
Given these caveats, the models provide an adequate fit. 
There are still some discrepancies present, 
such as the prominent main peak at [Fe/H]$_{CG}$
$\sim -1.7$~dex, which deviates by more than 2$\sigma$ (taken as $\sqrt{N}$) from 
the model curves and a single Gaussian. 
Note, however that this peak is not 
 distinct in the MDF on the ZW scale. 
As noted above, any potential substructure may reflect 
variations in star formation activity.  That such variations must have occurred is
obvious from the presence of multiple main-sequence turn-offs and the
decline in star formation activity in between.
Indeed, not just outflows
are implicated, but also inflows to sustain star formation over many Gyr
(see, e.g., Silk, Wyse \& Shields 1987, Lin \& Murray 1998; Carigi, Hernandez \& Gilmore 2002; 
Dong et al.\ 2003).  

\subsection{Infall and the gravitational potential}

Mac Low \& Ferrara (1999) investigated starburst-driven mass loss from
dwarf galaxies.  They distinguish between gas that is blown out but
retained by the gravitational potential well of their host galaxy, and gas
that is entirely blown away and lost from the galaxy.  Lin \& Murray (1998)
and Dong et al.\ (2003) present a model for episodic star formation in
dwarf galaxies like Carina according to which enriched gas is blown out but
eventually re-accreted once it has cooled down sufficiently, leading to a
new episode of star formation. Smecker-Hane et al.\ (1996) argued
that episodic accretion of fresh gas seemed unlikely since the mass of
Carina's dark halo seemed to be too low; also Carigi et al.\ (2002) note that in
the case of external accretion the previously measured tidal radius (Irwin
\& Hatzidimitriou 1995) needs to be larger, i.e., the potential well needs
to be deeper to permit the accretion of external gas.  

Odenkirchen et al.\ (2001) demonstrated that tidal radii inferred from
photographic data may underestimate the angular extent and nominal limiting
radius of dwarf galaxies.  For Carina we may expect a firmer limit on the
dark matter halo from the analysis of its radial velocity dispersion
profile (Wilkinson et al., in prep.).  Here we note that our spectroscopic
data indicate that four red giants whose radial velocities are consistent
with membership in Carina lie beyond of its nominal limiting stellar radius (Fig.\
11, top panel).  Three of these stars have metallicities consistent with
the trend seen in the outskirts of Carina.  Claims suggesting the existence
of ``extratidal'' stars around Carina have also been made by Kuhn, Smith,
\& Hawley (1996) and Majewski et al.\ (2000, 2005).  Whether this implies
that Carina is not in virial equilibrium and in the process of
losing/having lost substantial amounts of mass as suggested in these
studies is beyond the scope of the present paper.  However, if Carina were
a remnant of a once substantially more massive galaxy then the simple
chemical models discussed here would need to be modified. 

\subsection{A model with outflows and infall}

A more sophisticated model (Lanfranchi \& Matteucci 2004) is also plotted
in Fig.~13. This special model uses as input a SFH consistent with the CMD
and hence incorporates two major epochs of outflows and respective winds
associated with two main star formation episodes, each with a
duration of 3\,Gyr.  
The crucial model parameters were fit such as to
reproduce Carina's total stellar mass, the gas content and observed [$\alpha$/Fe]
abundance ratio for five stars.  The distribution resulting from this model
holds strictly for the true elemental iron abundance as measured from
high-resolution spectroscopy, rather than overall metallicity that we
derived from our CaT measurements (but see the discussion in Section 3.3).
The Lanfranchi \& Matteucci model for Carina shows one dominant metallicity
peak, demonstrating the insensitivity of the resulting MDF to the details
of the SFH, even for Fe.  There is again acceptable agreement between this
particular model (in a non-instantaneous recycling approximation) and our
observations, particularly when observational uncertainties are included
(right panel, Fig.~13).  
However, as before there are also remaining discrepancies, 
e.g., associated with the metal-rich and metal-poor tails 
(with a persisting G-dwarf problem). 

The Carina model of Lanfranchi \& Matteucci tries to reproduce individual
element abundance measurements for five stars and has hence constraints
only over the narrow metallicity range of these five stars.  Lanfranchi \&
Matteucci's ``best'' model (see Fig.\ 13) requires a wind efficiency of
some seven times the SFR and high star formation efficiency, leading to
solar metallicities within only a few Gyr.  This is at odds with our
MDF in combination with the photometrically
inferred ages of the different star formation episodes in Carina (e.g.,
Hurley-Keller et al.\ 1998; Monelli et al.\ 2003).  Lanfranchi \& Matteucci
point out that their results also differ from the SFH derived by Tolstoy et
al.\ (2003) on the basis of the five stars mentioned before.  Adding more
data points (Koch et al., in prep.) should aid in imposing stronger
constraints.  

The spatial variation of the SFH of the Carina dSph and other dSphs (e.g.,
Harbeck et al.\ 2001), plus evidence for cold sub-structures in dSphs
(e.g., Kleyna et al.~2004; Wilkinson et al.~2004) suggest that star
formation and chemical evolution should be modeled on small scales.  We
will present the details of such models in a future paper (Wyse et al., in
prep.).  

\section{Summary and Discussion}

Carina is the only dSph known to have undergone distinct, well-separated
episodic star formation.  The dominant episodes of star formation took
place at intermediate ages.  It is not yet understood what caused the
repeated cessation and delayed re-start of star formation in this enigmatic
dSph.  In the framework of a VLT Large Programme aiming exploring the
chemical evolution, kinematics, and dark matter content of Galactic dSphs,
we have derived CaT metallicities for 437 red giants in the Galactic dSph
Carina.  Our spectroscopic sample exceeds the previously largest published
dataset (Smecker-Hane et al.\ 1999) by a factor of $\sim 8$.  Our targets
cover the entire surface area of Carina and allow us to investigate also
radial metallicity variations.  

While we are using Ca as a tracer of overall metallicity (``[Fe/H]''), we
are well-aware of the different dependencies of [Fe/H] and [Ca/Fe] on the
detailed star-formation history of a galaxy, and of possible differences
between Ca and Fe abundances in globular clusters and dwarf galaxies.  We
compare our results to direct Fe measurements from the few existing
high-resolution spectra in the literature and arrive at the same
conclusions as previous authors:  The widely used CaT method is indeed a
powerful tool for efficient measurements of {\em overall}\ metallicities
and for studies of overall trends in the metallicity distribution of
intermediate-age and old populations in nearby galaxies as presented here.
However, uncertainties in the metallicity measurements for {\em individual} red
giants may amount to up to more than 
0.2 dex when using the CaT method.

The resulting metallicity distribution function shows a wide range of
metallicities spanning $\sim 3.0$ dex in [Fe/H].  The MDF peaks at a mean
metallicity of $-1.72\pm 0.01$\,dex (CG scale) or $-1.90\pm 0.01$ (ZW
scale), respectively.  Our mean metallicity on the CG scale is slightly
more metal-rich as compared to earlier spectroscopic CaT studies (e.g.,
Smecker-Hane et al.\ 1999).  The entire distribution's {\it formal} full
width at half maximum is 0.92\,dex.  
We proposed that some of this spread might  
be attributed to the occurrence of an extended SF and self-enrichment, 
resulting in several subpopulations with different
peak metallicities. 
The question whether these subpopulations of different metallicities 
are manifestations of corresponding distinct episodes of SF requires a combination
of deep photometric and spectroscopic data and detailled modelling
and will be left for a future paper. 
The MDF and its peaks change as one goes to larger
distances from Carina's center.  Generally, it shows a trend for more
pronounced metal-poor subpopulations with increasing distance.

We found a higher concentration of metal-rich, presumably intermediate-age
stars toward the inner regions of the galaxy as compared to more metal-poor
stars.  Metal-poor stars are, however, detected throughout the entire
galaxy.  This trend matches the observed central concentration of
intermediate-age stars in Carina (Harbeck et al.\ 2001), supporting that
the more metal-rich stars are also the younger ones.  Another argument in
support of an age-metallicity relation in Carina is the location of the
more metal-rich stars on the RGB:  Their metallicities in combination with
their photometric colors are inconsistent with these stars belonging to the
old population.  

Our data thus confirm earlier suggestions that Carina shows an
age-metallicity degeneracy in the sense that higher metallicities
counteract the effects of younger ages.  This conspiracy in turn leads to
the observed narrow RGB.  These findings and the considerable metallicity
spread rule out scenarios in which Carina's unusual episodic star formation history was
caused by the repeated accretion of pristine, unenriched gas. 

Comparing our MDFs to simple chemical evolution models such as a closed-box
model and prompt initial enrichment reveals a G-dwarf problem at low
metallicities, but a fairly close approximation of the shape of the MDF at
high metallicities. The metallicity peak of the MDF itself 
is not reproduced by these simple models.
One should bear 
in mind that the peak is most prominent on the CG scale, but does not show up in
the MDF based on the scale of ZW. 
The early star formation in Carina that
can still be traced by present-day red giants must have occurred from
pre-enriched gas (e.g., by Population III stars).  A better match to the
MDF can be obtained through more sophisticated models that include infall
and outflows.  The steep rise toward higher metallicities and its
culmination in a  prominent peak are qualitatively well reproduced by the
models of Lanfranchi \& Matteucci (2004), but the subsequent decline toward
even higher metallicities is too rapid compared to our observationally
derived MDF.  Still, the qualitatively good overall agreement between these
models and the observations is encouraging.  One possible explanation for
the repeated onset and cessation of the extended star formation episodes in
Carina is the possible re-accretion of previously blown-out, but not
blown-away, enriched gas (see, e.g., Lin \& Murray 1998; Dong et al.\
2003).  The predicted time scales for blow-out, cooling, and re-accretion
are consistent with the observed long pauses in star formation activity.  

Refined chemical evolution models will be presented in a forthcoming paper
(Wyse et al., in prep.), complemented by results from high-resolution
spectroscopy and a new derivation of the star formation history of Carina
that combines spectroscopic and photometric information (Koch et al., in
prep.).   This will also lead to a quantitative age-metallicity relation.
Furthermore, future papers will investigate whether different
subpopulations in Carina show different kinematics (Wilkinson et al., in prep.).

A handful of the red giants whose radial velocities make them likely Carina
members lie beyond the nominal tidal radius of Carina.  Spectroscopic
evidence for potential ``extratidal'' stars has also been found by Majewski
et al.\ (2000, 2005), raising the question whether Carina is a
dark-matter-dominated, bound dSph with an even larger limiting radius than
previously thought (see, e.g., Odenkirchen et al.\ 2001)  or a disrupted
remnant.  Carina's orbit as inferred from its proper motion indicates that
its perigalactic distance to the Milky Way is rather close ($\sim 20$ kpc),
suggesting that Carina needs to be strongly dark-matter dominated in order
to have survived these passages (Piatek et al.\ 2003).  The radial velocity
dispersion profiles that will be derived as part of of our VLT Large
Programme will help to elucidate the extent of Carina's dark matter halo
(Wilkinson et al., in prep.).

In spite of its unusual star formation history, Carina fits in well with
the metallicity-luminosity relation of other dSph galaxies. The
metallicity-luminosity relation of dSphs is offset from that of dIrrs
such that dSphs have higher mean stellar metallicities for a fixed optical
luminosity.  This offset suggests that star formation in young dSphs was
more vigorous than in young dIrrs (Grebel et al.\ 2003).  As is the case
for other dSphs, Carina is thus unlikely to simply be a stripped former
irregular galaxy.  However, Piatek et al.\ suggest that Carina is currently
near apogalacticon and that its pericentric passages bring it very close to
the Milky Way, while its star formation history does not seem to be
correlated with its orbital motion.  Nonetheless, Carina's orbit would seem
to suggest that in addition to intrinsic effects such as feedback, its
evolution must also have been influenced by externally induced mass loss.
Such mass loss may have been induced through tidal interactions with the
Milky Way and/or through ram pressure stripping when passing through the
denser regions of the Galaxy's gaseous halo.  
There appears to be a correlation between present-day Galactocentric distance and the
fraction of intermediate-age populations (van den Bergh 1994; Grebel 1997,
1999). Carina with its distance of 94\,kpc and its prominent intermediate-age population seems 
to fit into this correlation rather well. However, one may ask whether this is merely fortuitous 
for this galaxy,if one considers Carina's proposed
short pericentric distance and the resulting eccentric orbit. 
 Clearly, a comprehensive
exploration of Carina's history will have to take into account both
internal and external influences, its detailed chemical evolution, its age
structure, its mass and mass distribution, its internal kinematics and its
orbit.

\acknowledgments

We thank Andrew Cole for providing his version of the Equivalent Width
code,  Gustavo Lanfranchi for sending us the output of the Carina model, 
and Michael Odenkirchen for his help with the fiber
configurations for the first runs of this program.
AK and EKG gratefully acknowledge support by the Swiss National
Science Foundation through grant 200021-101924 and 200020-105260.   
MIW acknowledges PPARC for financial support.

\clearpage

\begin{figure} 
\includegraphics[width=12cm]{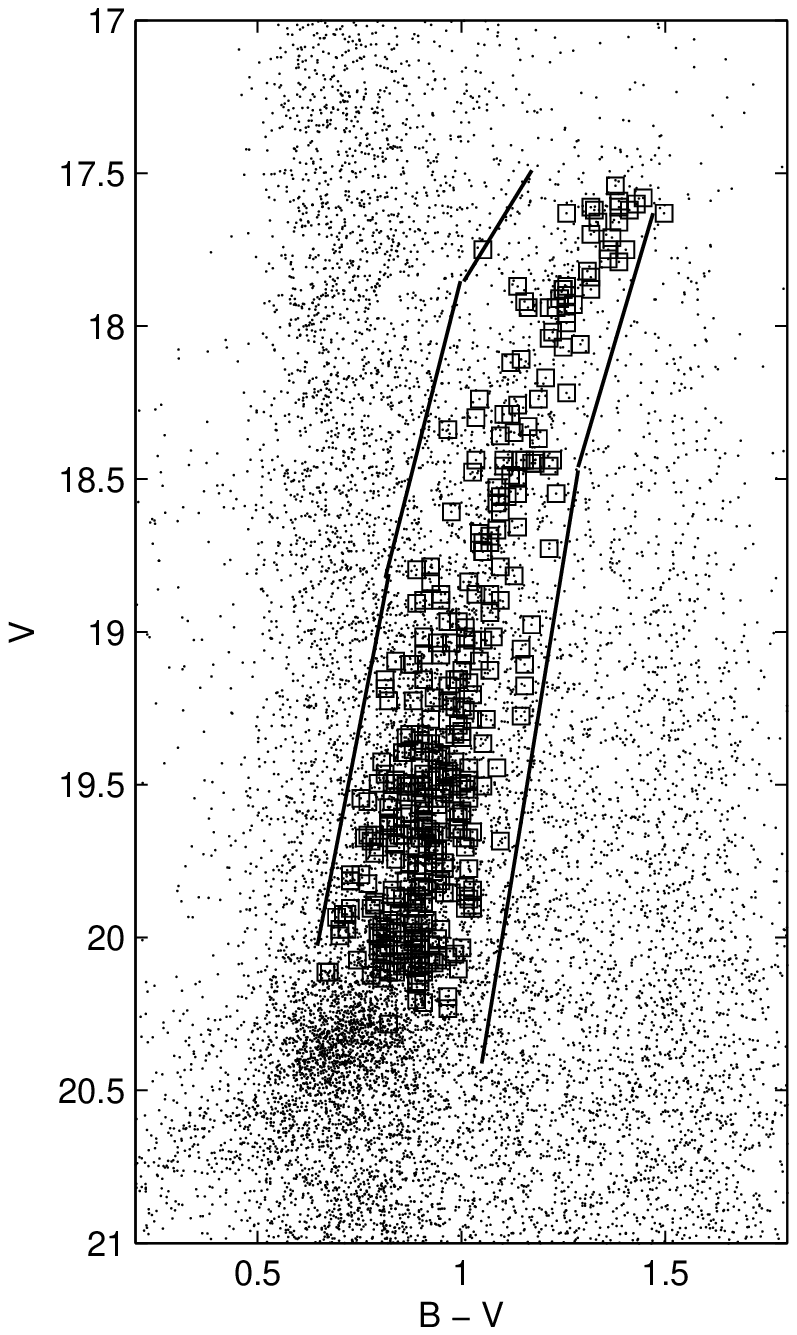} 
\caption{
Region of the red giant branch in Carina, from which our targets were
selected.  The photometry shown here is EIS photometry transformed into the
Landolt-Stetson {\em UBVRI} system. The solid lines delineate the color cuts imposed to select the targets
and open squares denote the confirmed
radial velocity members of Carina.} 
\end{figure}

\begin{figure}
\plotone{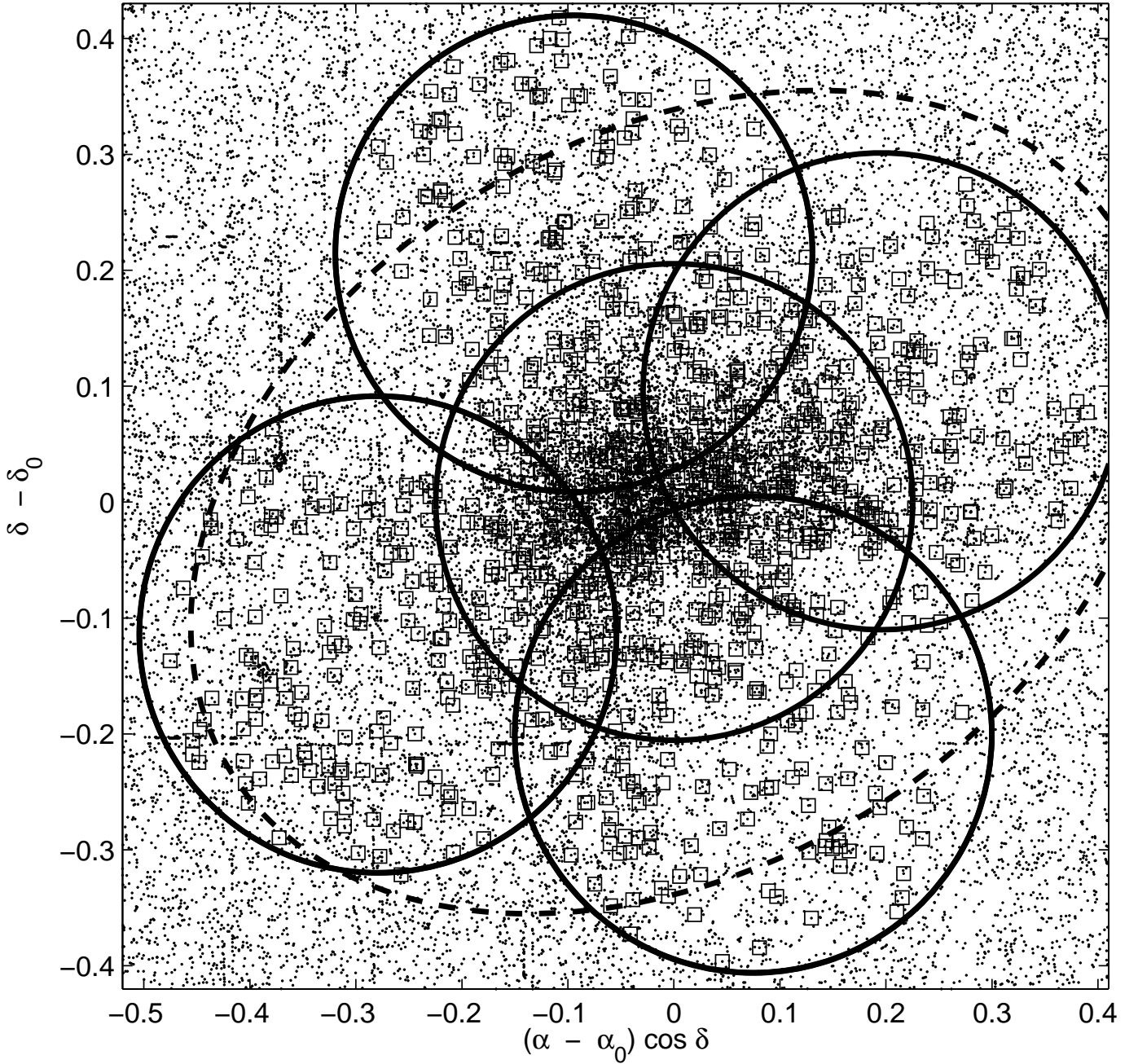}
\caption{
Location of our targets (shown as open squares) relative to Carina's 
geometric centre ($\alpha_0$,$\delta_0$), superposed on a stellar map
of the galaxy and its surroundings. The circles circumscribe the field
of view of the FLAMES spectrograph of 25$\arcmin$.  The ellipse indicates
the nominal tidal radius, ellipticity, and position angle of Carina as
inferred by Irwin \& Hatzidimitriou (1995).}
\end{figure}

\begin{figure}
\plotone{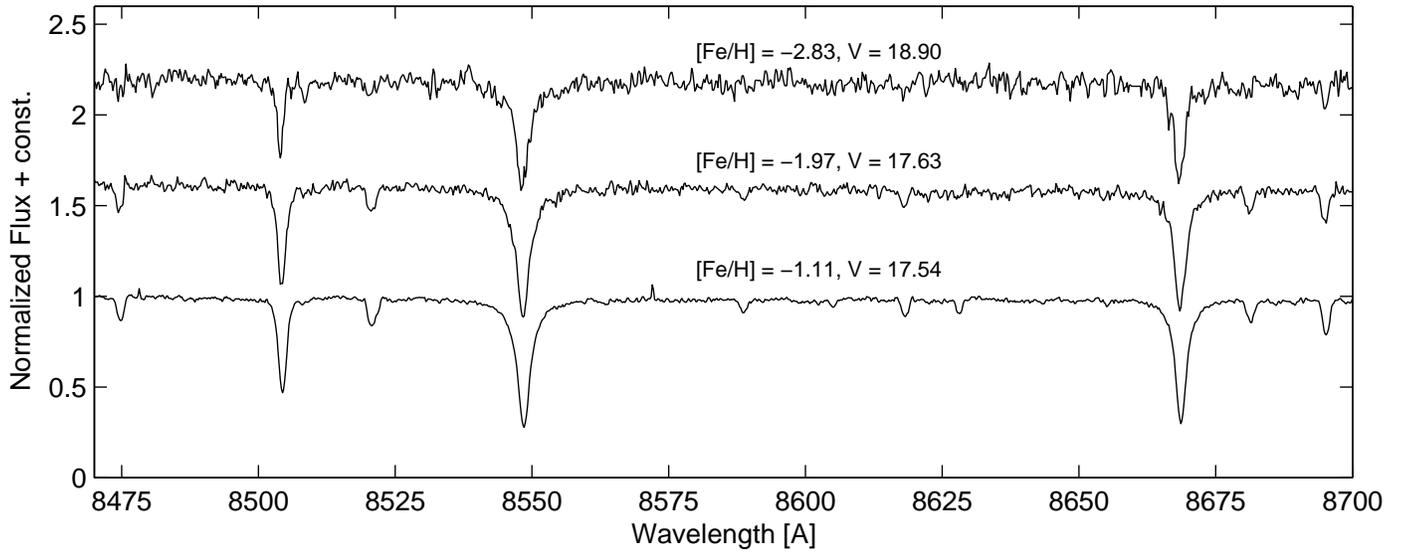}
\caption{Sample spectra of Carina red giants centered on the near-infrared
Ca\,{\sc ii} triplet region.  The three spectra are representative of the 
observed range in metallicities.}
\end{figure}

\begin{figure}
\plotone{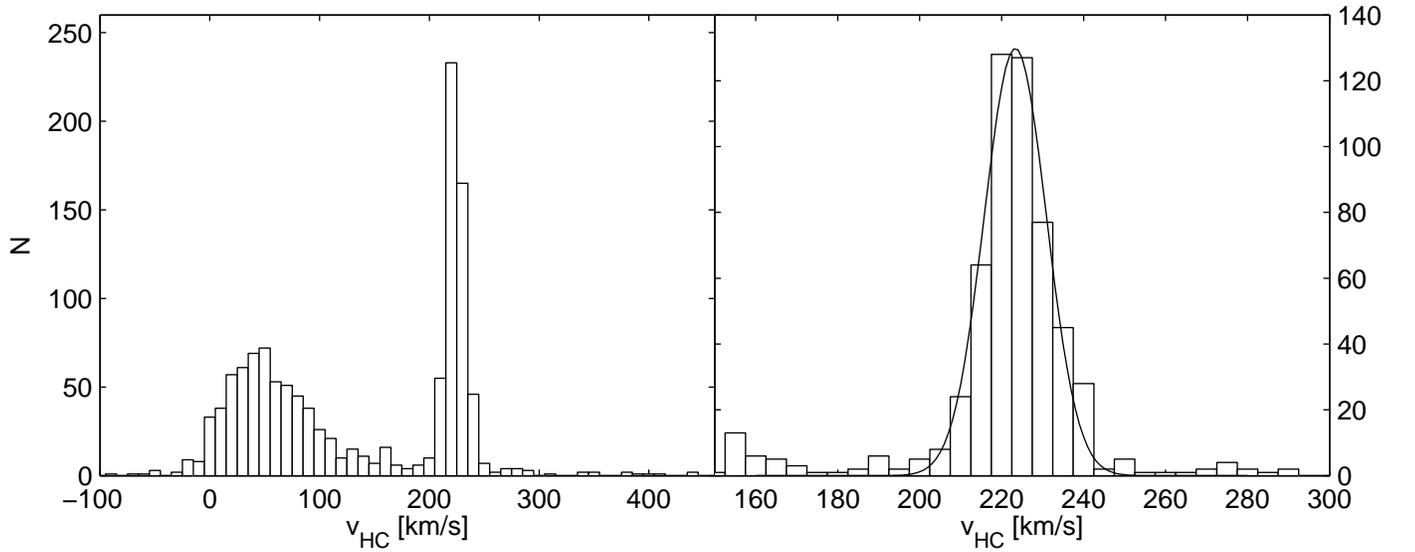}
\caption{Left panel: Distribution of radial velocities for all stars 
in our sample.  Right panel: The region around Carina's systemic 
velocity of $\sim$223\,km\,s$^{-1}$.  Also indicated is the best fit 
of a single Gaussian.}
\end{figure}

\begin{figure}
\plotone{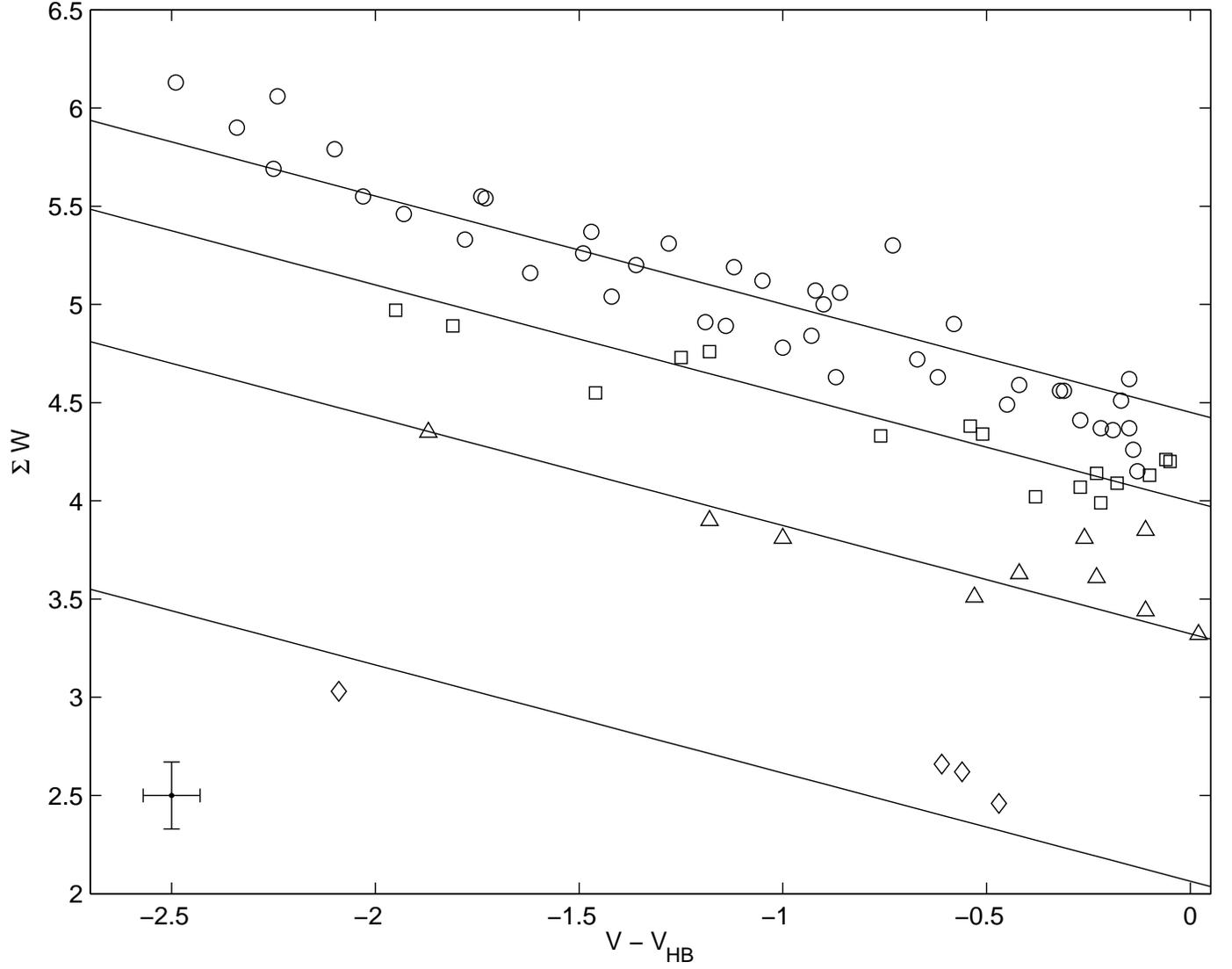}
\caption{Equivalent widths of red giant branch stars in our four
calibrating globular clusters plotted against their magnitude above their
respective horizontal branches. Shown as a solid line is the best-fitting 
line for each cluster, based on a common slope of 0.55\AA\,mag$^{-1}$. 
The symbols denote NGC\,5904 (open circles), NGC\,3201 (open squares), 
NGC\,4147 (open triangles) and NGC\,4590 (open diamonds). A 
representative 1$\sigma$ errorbar is indicated at the lower left.
The metallicities of the globular clusters decrease from top to bottom.
See Table~1 for their [Fe/H] values on different metallicity scales.}
\end{figure}

\begin{figure}
\plotone{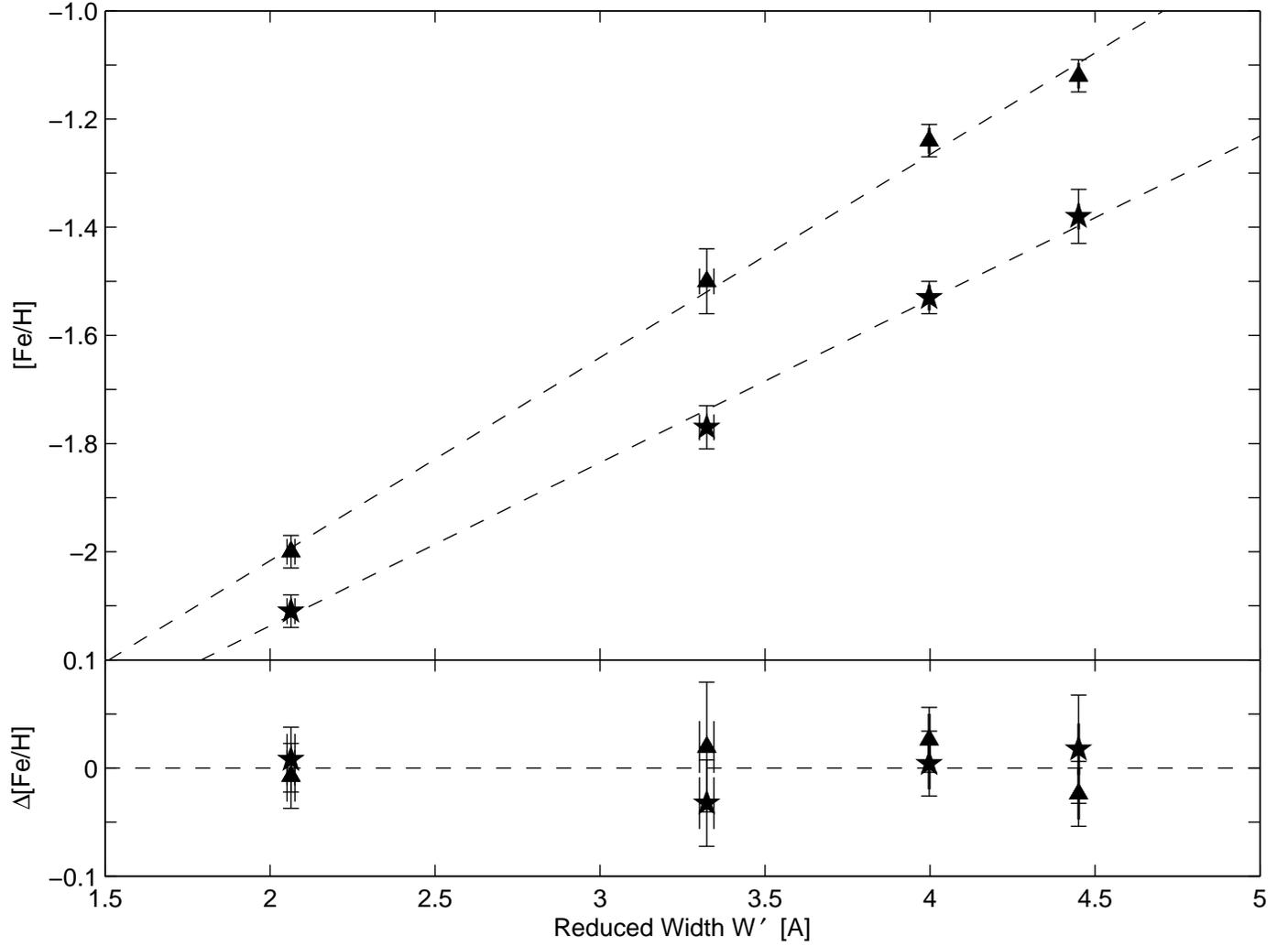}
\caption{The top panel displays the metallicity of the four calibration 
clusters against their reduced width, both calibrated against the 
reference values of Rutledge et al.\ (1997b) on the scale of Zinn \& 
West (stars) and Carretta \& Gratton (triangles).  The residuals of the 
linear best fit (dashed lines) are plotted in the bottom panel.}
\end{figure}

\begin{figure}
\plotone{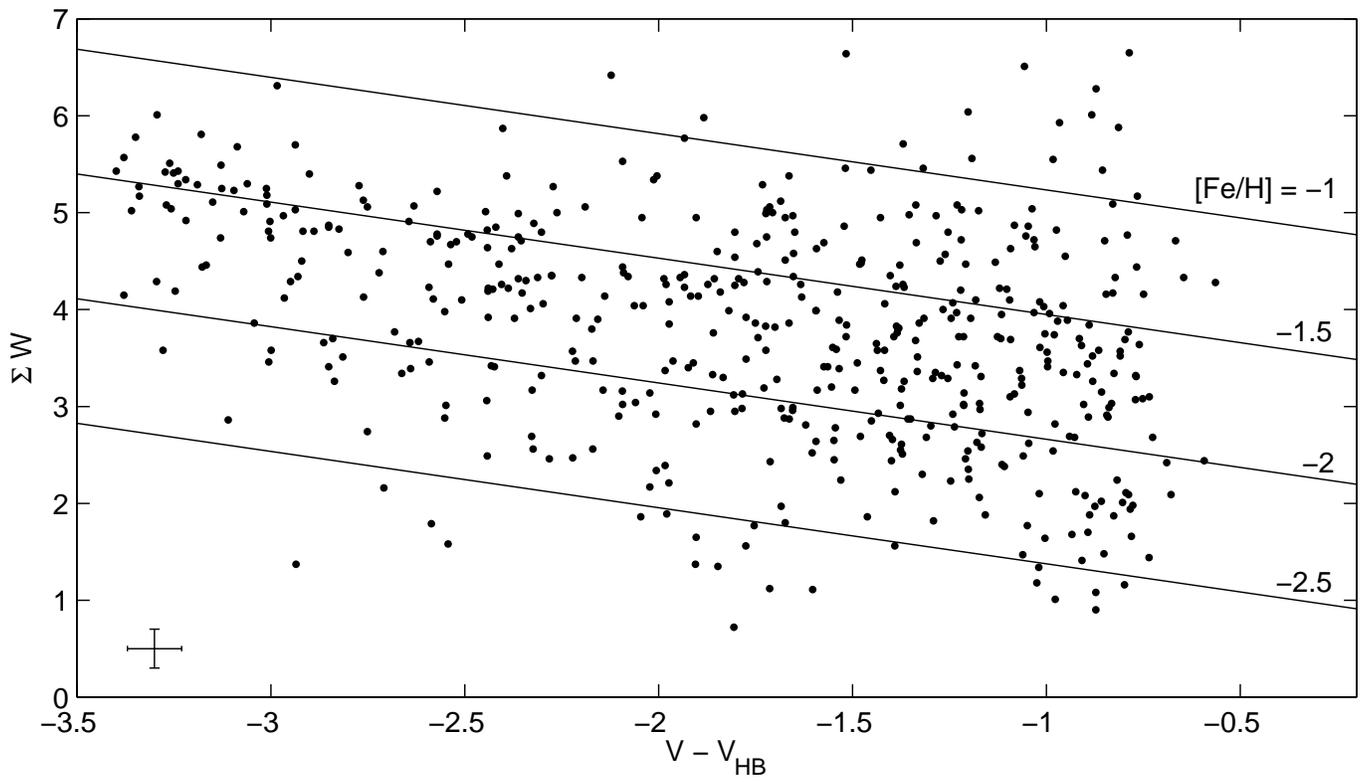}
\caption{Distribution of the Carina targets in the V,$\Sigma$W-plane. 
As the lines of constant metallicity indicate, the galaxy exhibits a 
large spread in [Fe/H].  A typical errorbar is shown left.}
\end{figure}

\begin{figure}
\plotone{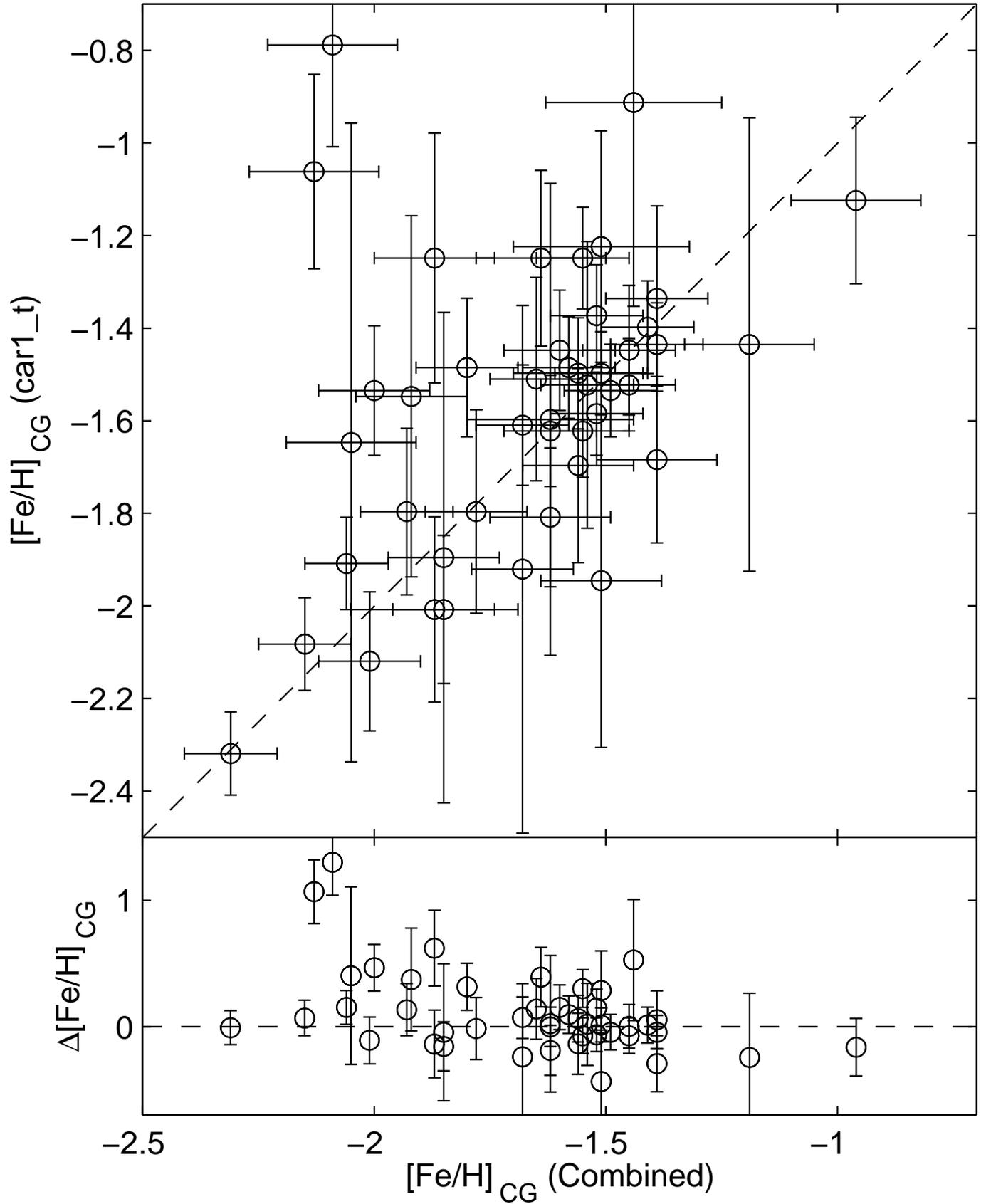}
\caption{Comparison of the stars observed during different observing 
runs: Shown are metallicities derived from spectra of the first run 
(labeled {\em car1\_t}) versus [Fe/H] as measured from the combined 
spectra of all runs. The dashed line is unity and the bottom panel 
displays the respective residuals.}
\end{figure}

\begin{figure}
\plotone{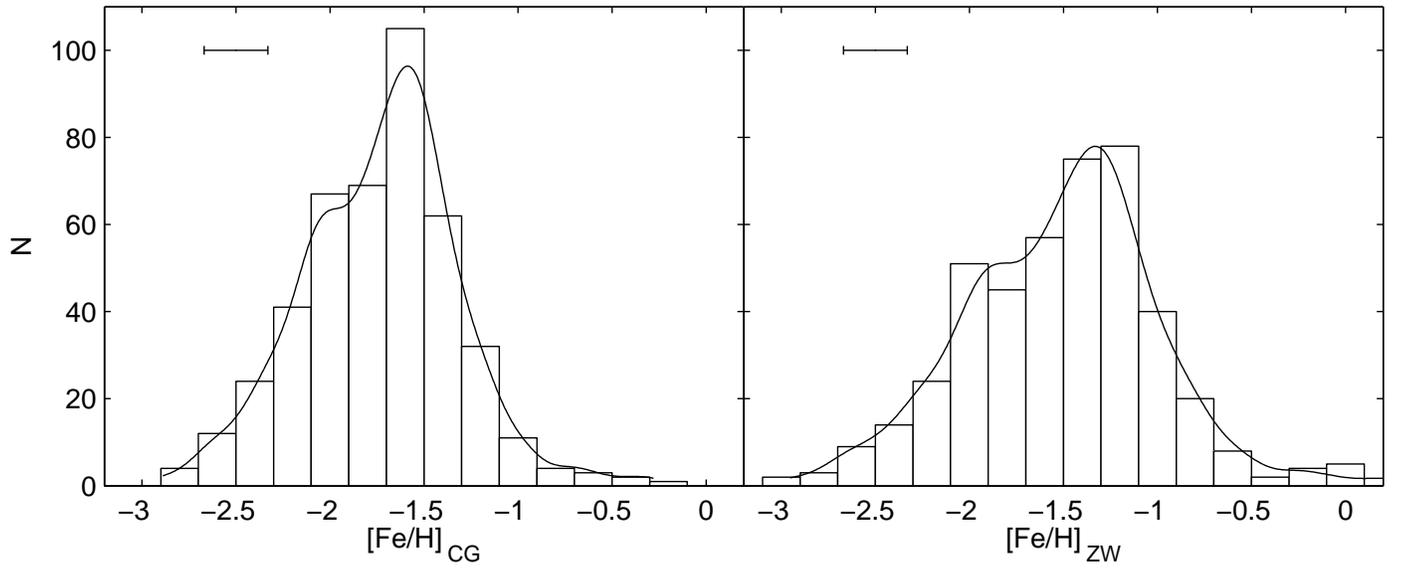}
\caption{Histograms for metallicities of the RGB stars in Carina. The 
histogram in the left panel uses the scale of CG,  the right one refers 
to the scale of ZW.  The line at the top right indicates the median 
measurement error of 0.17\,dex. Overplotted as solid lines are MDFs, convolved by 
the respective errors.}
\end{figure}

\begin{figure}
\plotone{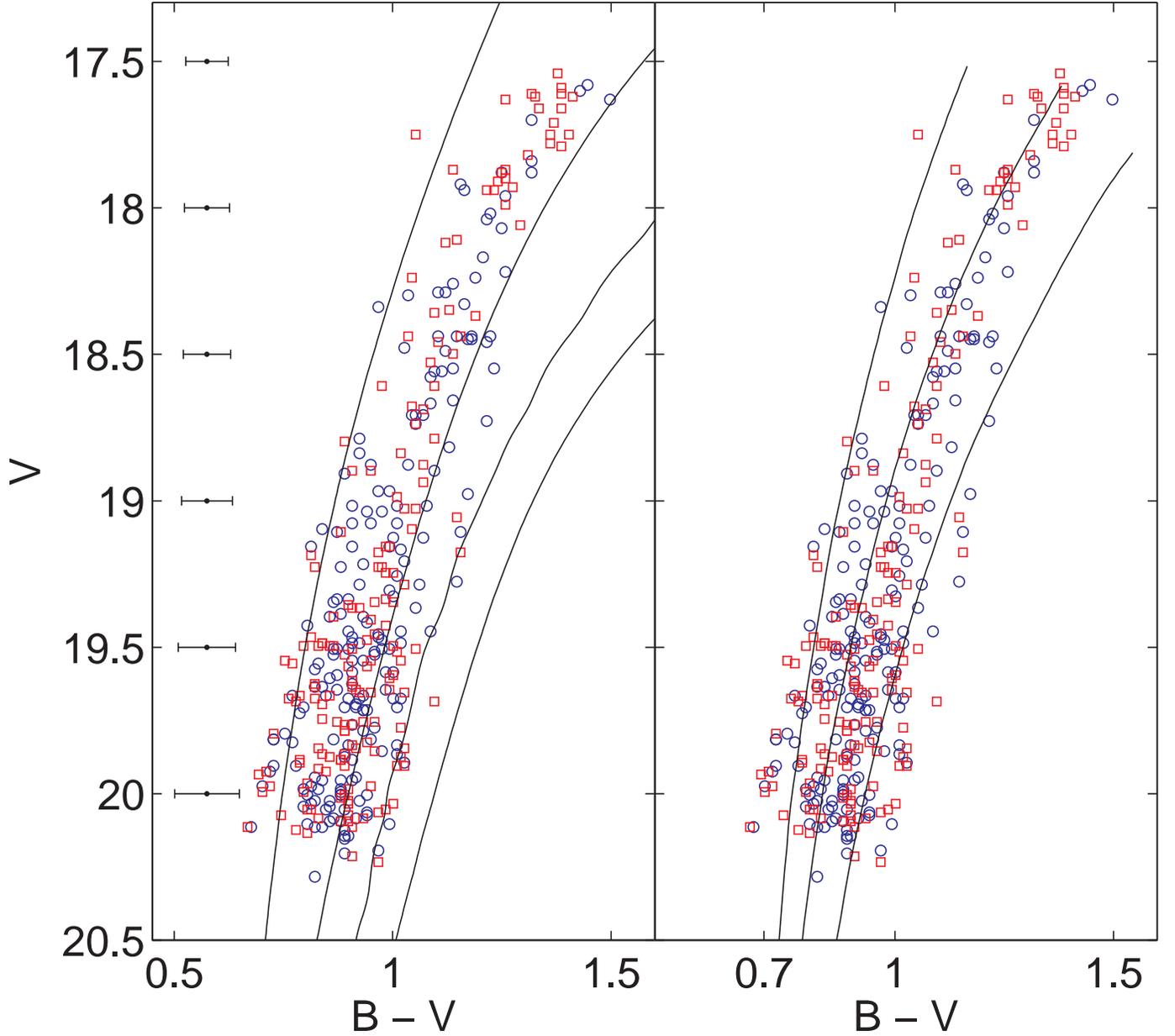}
\caption{
Left panel:  Distribution of metallicities across the RGB.  Red squares 
depict the metal-poor RGB stars with [Fe/H]$_{\mathrm{CG}}\le -$1.68, 
whereas blue circles refer to the metal-rich tail of the distribution 
([Fe/H]$_{\mathrm{CG}}>-$1.68).  Also indicated as solid lines are 
globular cluster fiducials from Sarajedini \& Layden (1997) at the 
respective [Fe/H]$_{\mathrm{CG}}$, from left to right: M15 ($-2.02$), 
NGC\,6752 ($-1.24$), NGC\,1851 ($-1.03$) and 47\,Tuc ($-0.78$).
Representative color errorbars are indicated in the left panel. 
--- Right panel:  The same data points shown with Yonsei-Yale isochrones
(solid lines; Yi, Kim, \& Demarque 2003).  {From} left to right, the
isochrones have the following parameters: [Fe/H] = $-2.3$ dex and 12.6 Gyr,
$-1.7$ dex and 6.3 Gyr, $-1.3$ dex and 3.2 Gyr.  The choice of parameters
corresponds roughly to the three main episodes of an assumed star formation history
of Carina.  Comparing the location of the isochrones with the red giant
branch fiducials on the left-hand side illustrates the effects of the
age-metallicity degeneracy. }
\end{figure}

\begin{figure}
\plotone{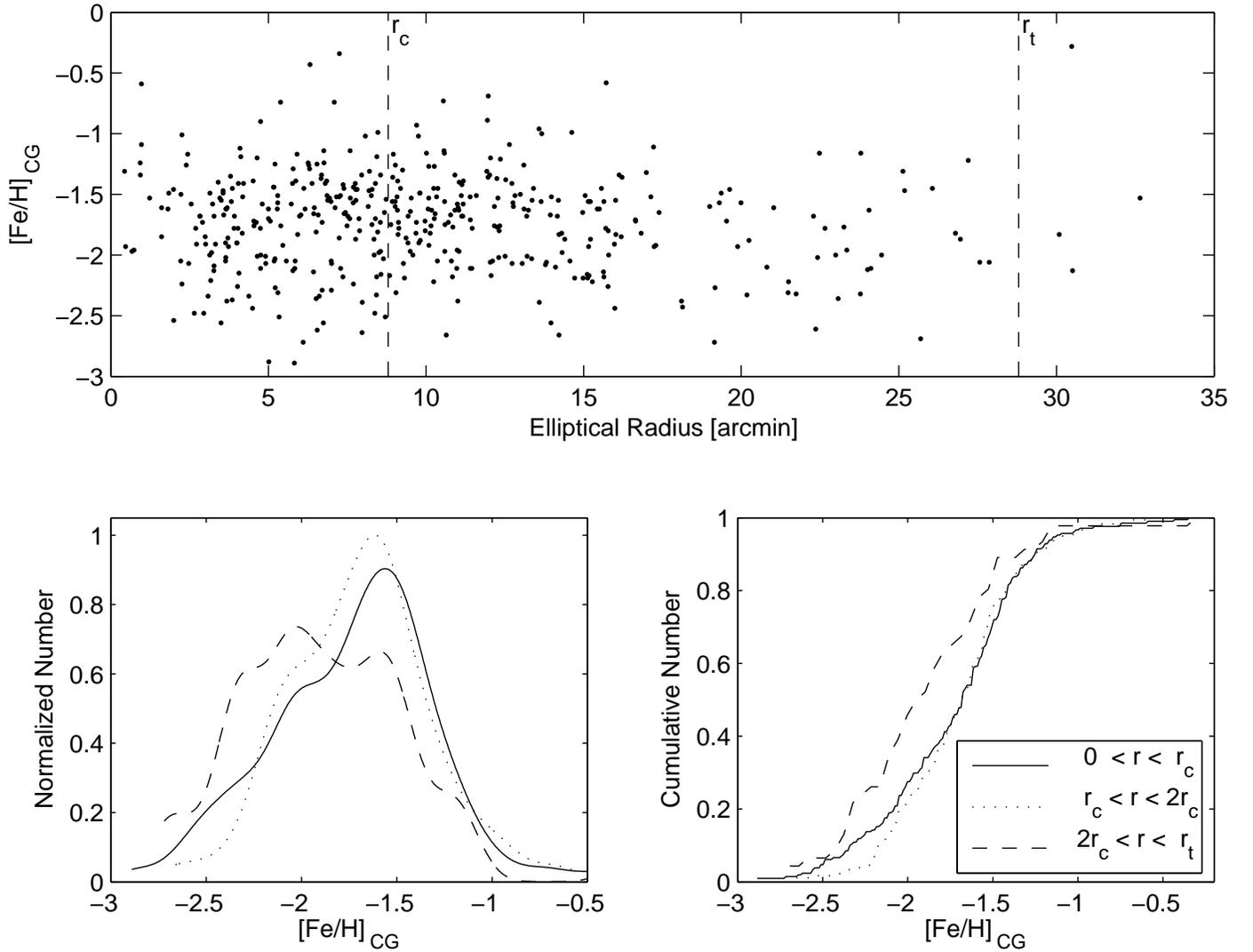}
\caption{CaT metallicities of the 437 red giants versus their elliptical 
radius in Carina (top). Nominal core and tidal radii are also denoted. 
The bottom left panel shows MDFs for three regions at different radii. 
These density distributions were convolved by the individual measurement 
errors in [Fe/H]. A weak trend of the MDF to become more metal-poor when 
proceeding outwards is visible and also reflected in the 
cumulative distribution (bottom right).}
\end{figure}
\begin{figure}
\plotone{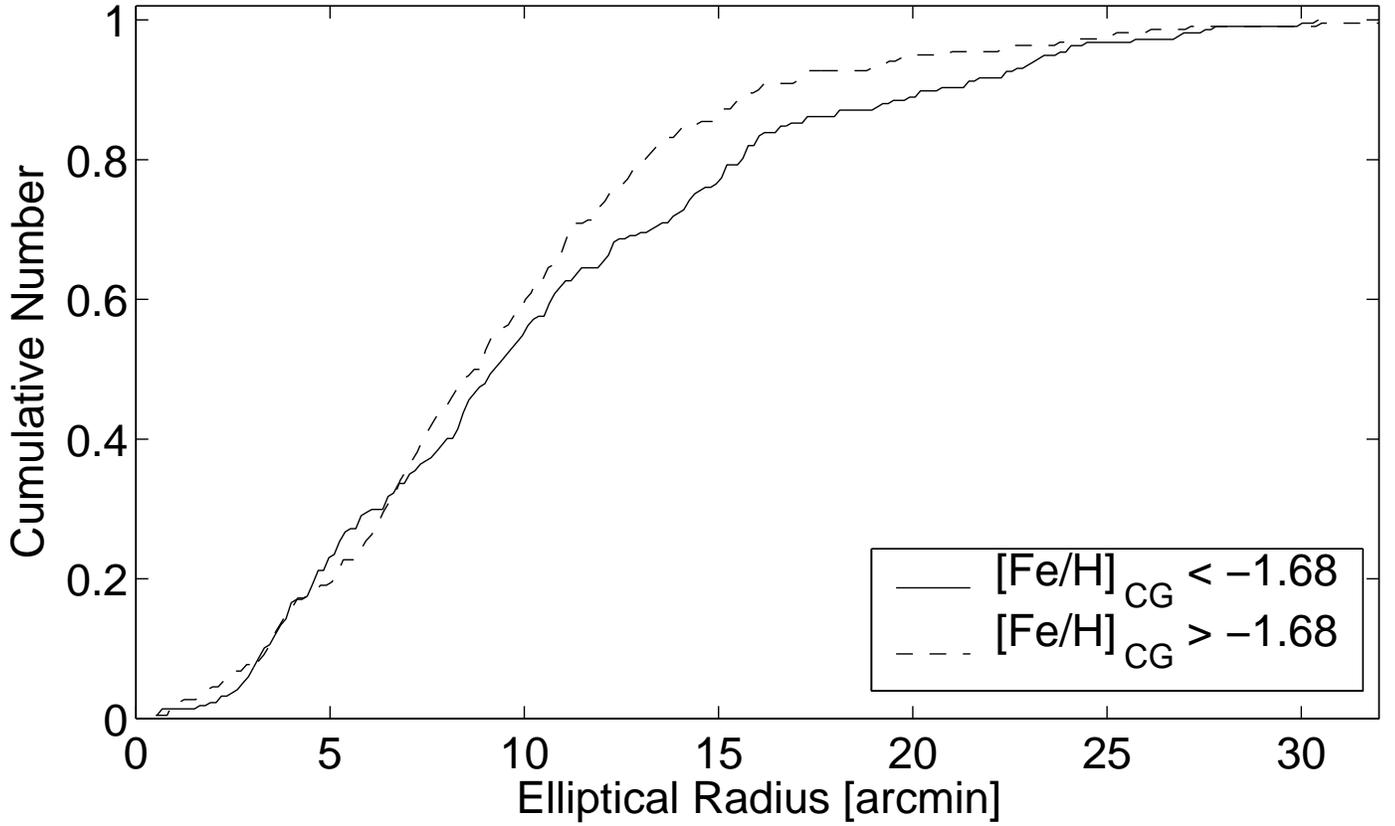}
\caption{Cumulative number distributions for elliptical radii of the 
metal-poor population ([Fe/H]$<-1.68$, solid line) and the more 
metal-rich component (dashed line). 
The latter is more centrally concentrated.}
\end{figure}
\begin{figure}
\plotone{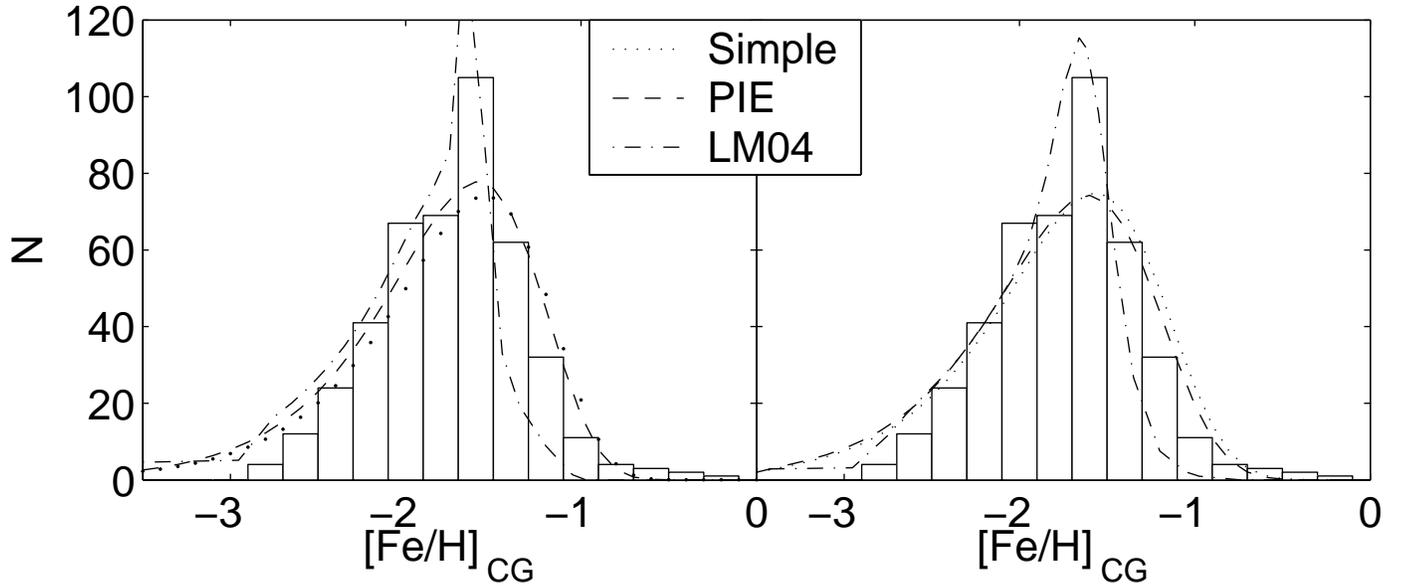}
\caption{Fits of basic models (
modified simple closed box 
with outflows
and prompt initial 
enrichment (PIE)) overplotted on the frequency distribution of the 
observed metallicities, where the dash-dotted line refers to 
the predicted distribution for Carina from Lanfranchi \& Matteucci 
(2004, LM04), scaled to match the total number of stars. The right panel 
is the analogous plot where the models were convolved by the observational
errors.}
\end{figure}

\clearpage

\begin{sidewaystable*}
\begin{center}
\caption{Globular Clusters used for calibration of the metallicity scale.}
\begin{tabular}{cccccccccc}
\hline
\hline
Cluster & $\alpha$ (J2000) & $\delta$ (J2000) & [Fe/H]$_{ZW,R97}$ & [Fe/H]$_{CG,R97}$ 
& (m$-$M)$_V$ & E\,(\bv) & V$_{HB}$ & [Ca/Fe] & References \\ 
\hline
NGC\,3201 & 10 17 36.8 & $-$46 24 40 & $-$1.53\,$\pm\,$0.03 & $-1.24 \pm 0.03$ & 14.21 & 0.21 & 14.74\,$\pm\,0.07$& 0.11 & 1,2,3,4 \\
NGC\,4147 & 12 10 06.2 &   +18 32 31 & $-$1.77\,$\pm\,$0.04 & $-1.50 \pm 0.06$ & 16.48 & 0.02 & 16.95\,$\pm\,0.10$ & $\dots$ & 1,2,3 \\
NGC\,4590 & 12 39 28.0 & $-$26 44 34 & $-$2.11\,$\pm\,$0.04 & $-2.00 \pm 0.03$ & 15.19 & 0.04 & 15.75\,$\pm\,0.05$ & 0.32 & 1,2,3,4 \\
NGC\,5904 & 15 18 33.8 &   +02 04 58 & $-$1.38\,$\pm\,$0.05 & $-1.12 \pm 0.03$ & 14.46 & 0.03 & 15.13\,$\pm\,0.05$ & 0.21 & 1,2,3,4 \\
\hline
&&&&&&&&&\\
\multicolumn{10}{l}{References. ---  (1) Harris (1996); (2) Rutledge et al. (1997); (3) Ferraro et al. (1999); (4) Carney (1996)} \\
\end{tabular}
\end{center}
\end{sidewaystable*}

\begin{table}
\begin{center}
\caption{Observation log}
\begin{tabular}{clc}
\hline
\hline
Date & Field (Configuration) & Total exp. time [s] \\
\hline
21 Feb 2003 &  Center  &1560\\
 	   & NGC\,3201  &627\\
22 Feb 2003 &  Center  &9900\\
           & NGC\,4147 & 2400\\
03 Mar 2003 &  Center  &9497\\
04 Mar 2003 &  Center & 7800\\
	   & NGC\,5904  &3600\\
05 Mar 2003 &  Center  &7800\\
	  &  NGC\,4590 & 1800\\
	  &  NGC\,3201  &1200\\
22 Dec 2003 &  NE (1c) & 14981\\
          &  NE (1b)  &12975\\
23 Dec 2003 &  Center (2b)  &13630\\
          &  Center (2a) &13213\\
24 Dec 2003 &  SW (1e)  &14401\\
	  &  SW (1d)  &13172\\
25 Dec 2003 &  NW (1e) & 14333\\
	  &  NW (1d)  &13113\\
26 Dec 2003 &  NE (2b)  &14470\\
	  &  NE (2a)  &13001\\
27 Dec 2003 &  SE (1e)  &20488\\
	  &  SE (1d) & 13333\\
28 Dec 2003 &  SW (2a)  &19469\\
	  &  SW (2b) & 235\\
	  &  NW (1e) & 7933\\
29 Dec 2003 &  SW (2b) & 13421\\
 	  &  Center (1e) & 13553\\
30 Dec 2003 &  Center (1d)  &13694\\
	  &  Center (2a) & 13393\\
21 Feb 2004 &SE (2a) & 19677\\
22 Feb 2004 &SE (2b)  &15162\\
	  &  NGC\,4590  &4794\\
	  &  NGC\,3201  &3598\\
23 Feb 2004 &NW (2a)  &10824\\
	  &  NGC\,4590 & 6596\\
24 Feb 2004 &NW (2b) & 10833\\
	  &  NGC\,3201 & 7196\\
25 Feb 2004 & NE (1d) & 23481\\
26 Feb 2004 & NE (1e) & 10884\\
27 Feb 2004 & NE (2a) & 10931\\
28 Feb 2004 & NE (2b) & 10635\\
\hline
&&\\
\end{tabular}
\end{center}
\hspace{4.5cm}\begin{minipage}{12cm}{
{\sc note} --- Labels in brackets refer to different configurations\\on the fiber positioning plates, 
thus aiming at different targets.\\ 
Only nights relevant for Carina observations are listed.}\end{minipage}
\end{table}

\begin{table}
\begin{center}
\caption{Observed fields in Carina and the calibration clusters}
\begin{tabular}{lcc}
\hline
\hline
Field & $\alpha$ (J2000) & $\delta$ (J2000)\\
\hline
Center & 06 41 36.8 & $-$50 57 58  \\
NW & 06 41 00.8 & $-$50 45 06 \\
NE & 06 42 51.1 & $-$50 52 50 \\
SW & 06 39 50.0 & $-$51 04 49 \\
SE & 06 42 05.3 & $-$51 10 00 \\
NGC\,3201 & 10 17 36.8 & $-$46 24 40\\
NGC\,4147 & 12 10 09.6 & +18 35 511\\
NGC\,4590 & 12 39 28.0 & $-$26 44 33\\
NGC\,5904 & 15 18 03.5 & +02 06 41\\
\hline
\end{tabular}
\end{center}
\end{table}

\begin{table}
\begin{center}
\caption{Stars in common with Shetrone et al.~(2003).}
\begin{tabular}{cccccc}
\hline
\hline
Star ID      &                                  &  [Ca/Fe] & [Fe/H]$_{\mathrm{high\,res}}$ & [Fe/H]$_{\mathrm{CaT,ZW}}$ & [Fe/H]$_{\mathrm{CaT,CG}}$ \\
(Mateo 1993) & \raisebox{1.5ex}[-1.5ex]{EIS-ID} & \multicolumn{2}{c}{(Shetrone et al. 2003)} &  \multicolumn{2}{c}{(This work)} \\
\hline
M2   & LG04a\_003764 & $+0.20 \pm 0.05$ & $-1.60 \pm 0.02$ & $-1.67 \pm 0.09 $  & $-1.43 \pm 0.09 $\\
M3   & LG04a\_001419 & $-0.10 \pm 0.06$ & $-1.65 \pm 0.02$ & $-2.00 \pm 0.08 $  & $-1.84 \pm 0.08 $\\
M4   & LG04a\_001673 & $+0.14 \pm 0.04$ & $-1.59 \pm 0.02$ & $-1.65 \pm 0.09 $  & $-1.41 \pm 0.09 $\\
M10  & LG04d\_006644 & $-0.02 \pm 0.05$ & $-1.94 \pm 0.02$ & $-2.16 \pm 0.08 $  & $-2.04 \pm 0.08 $\\
M12  & LG04a\_007126 & $+0.12 \pm 0.05$ & $-1.40 \pm 0.02$ & $-1.33 \pm 0.11 $  & $-1.01 \pm 0.11 $\\
\hline
\end{tabular}
\end{center}
\hspace{3cm}\begin{minipage}{12cm}{{\sc note} --- 
Metallicities given for the Shetrone et al.~(2003) stars are the weighted means from [Fe\,I/H] and [Fe\,II/H].}
\end{minipage}
\end{table}

\begin{table}
\begin{center}
\caption{Stars measured in more than one observing runs.}
\begin{tabular}{cccrc}
\hline
\hline
Star & [Fe/H]$_{CG}$, first run &  [Fe/H]$_{CG}$, second run &  $\Delta$[Fe/H]$_{CG}$& [Fe/H]$_{CG}$, combined spectra\\
\hline
LG04a\_000030 & $ -1.49 \pm 0.11$ & $ -1.77 \pm 0.10$ & $  0.28  $ & $-1.58 \pm 0.10$ \\
LG04a\_000377 & $ -1.45 \pm 0.14$ & $ -1.45 \pm 0.09$ & $  0.00  $ & $-1.45 \pm 0.10$ \\
LG04a\_000451 & $ -1.95 \pm 0.36$ & $ -1.45 \pm 0.12$ & $ -0.50  $ & $-1.51 \pm 0.13$ \\
LG04a\_001042 & $ -1.53 \pm 0.10$ & $ -1.36 \pm 0.10$ & $ -0.17  $ & $-1.49 \pm 0.10$ \\
LG04a\_001111 & $ -1.60 \pm 0.51$ & $ -1.75 \pm 0.18$ & $  0.15  $ & $-1.62 \pm 0.18$ \\
LG04a\_001142 & $ -2.01 \pm 0.16$ & $ -1.61 \pm 0.11$ & $ -0.40  $ & $-1.85 \pm 0.11$ \\
LG04a\_001170 & $ -1.62 \pm 0.12$ & $ -1.53 \pm 0.13$ & $ -0.09  $ & $-1.62 \pm 0.10$ \\
LG04a\_001234 & $ -0.91 \pm 0.44$ & $ -1.53 \pm 0.17$ & $  0.62  $ & $-1.44 \pm 0.19$ \\
LG04a\_001298 & $ -0.79 \pm 0.32$ & $ -2.32 \pm 0.19$ & $  1.53  $ & $-2.09 \pm 0.18$ \\
LG04a\_001313 & $ -1.06 \pm 0.21$ & $ -2.24 \pm 0.14$ & $  1.18  $ & $-2.13 \pm 0.14$ \\
LG04a\_001364 & $ -1.61 \pm 0.13$ & $ -1.57 \pm 0.09$ & $ -0.04  $ & $-1.68 \pm 0.10$ \\
LG04a\_001390 & $ -1.68 \pm 0.18$ & $ -1.62 \pm 0.21$ & $ -0.06  $ & $-1.39 \pm 0.13$ \\
LG04a\_001426 & $ -1.92 \pm 0.57$ & $ -1.58 \pm 0.11$ & $ -0.34  $ & $-1.68 \pm 0.11$ \\
LG04a\_001556 & $ -1.40 \pm 0.10$ & $ -1.41 \pm 0.09$ & $  0.01  $ & $-1.41 \pm 0.10$ \\
LG04a\_001558 & $ -1.81 \pm 0.15$ & $ -1.40 \pm 0.10$ & $ -0.41  $ & $-1.62 \pm 0.13$ \\
LG04a\_001608 & $ -1.49 \pm 0.15$ & $ -1.67 \pm 0.09$ & $  0.18  $ & $-1.80 \pm 0.11$ \\
LG04a\_001734 & $ -1.80 \pm 0.22$ & $ -1.93 \pm 0.10$ & $  0.13  $ & $-1.78 \pm 0.11$ \\
LG04a\_001875 & $ -1.53 \pm 0.14$ & $ -2.48 \pm 0.15$ & $  0.95  $ & $-2.00 \pm 0.12$ \\
LG04a\_001899 & $ -1.55 \pm 0.39$ & $ -2.01 \pm 0.12$ & $  0.46  $ & $-1.92 \pm 0.12$ \\
LG04a\_001910 & $ -1.50 \pm 0.12$ & $ -1.61 \pm 0.15$ & $  0.11  $ & $-1.56 \pm 0.14$ \\
LG04a\_001917 & $ -1.45 \pm 0.13$ & $ -1.66 \pm 0.13$ & $  0.21  $ & $-1.60 \pm 0.12$ \\
LG04a\_002065 & $ -1.37 \pm 0.11$ & $ -1.55 \pm 0.09$ & $  0.18  $ & $-1.52 \pm 0.10$ \\
LG04a\_002169 & $ -1.50 \pm 0.09$ & $ -1.52 \pm 0.09$ & $  0.02  $ & $-1.51 \pm 0.10$ \\
LG04a\_002181 & $ -1.44 \pm 0.09$ & $ -1.37 \pm 0.09$ & $ -0.07  $ & $-1.39 \pm 0.10$ \\
LG04a\_003952 & $ -1.25 \pm 0.19$ & $ -1.68 \pm 0.11$ & $  0.43  $ & $-1.64 \pm 0.14$ \\
LG04a\_004179 & $ -1.80 \pm 0.18$ & $ -1.92 \pm 0.09$ & $  0.12  $ & $-1.93 \pm 0.10$ \\
LG04b\_002633 & $ -1.25 \pm 0.27$ & $ -2.10 \pm 0.11$ & $  0.85  $ & $-1.87 \pm 0.13$ \\
LG04b\_005294 & $ -1.22 \pm 0.25$ & $ -1.90 \pm 0.25$ & $  0.68  $ & $-1.51 \pm 0.19$ \\
LG04c\_000626 & $ -1.91 \pm 0.10$ & $ -2.11 \pm 0.08$ & $  0.20  $ & $-2.06 \pm 0.09$ \\
LG04c\_000710 & $ -1.44 \pm 0.49$ & $ -1.24 \pm 0.15$ & $ -0.20  $ & $-1.19 \pm 0.14$ \\
LG04c\_003085 & $ -1.12 \pm 0.18$ & $ -1.16 \pm 0.11$ & $  0.04  $ & $-0.96 \pm 0.14$ \\
LG04c\_004153 & $ -1.70 \pm 0.21$ & $ -1.61 \pm 0.20$ & $ -0.09  $ & $-1.56 \pm 0.12$ \\
LG04c\_004227 & $ -2.32 \pm 0.09$ & $ -2.53 \pm 0.10$ & $  0.21  $ & $-2.31 \pm 0.10$ \\
LG04c\_004308 & $ -2.08 \pm 0.10$ & $ -2.13 \pm 0.08$ & $  0.05  $ & $-2.15 \pm 0.10$ \\
LG04c\_006477 & $ -1.25 \pm 0.11$ & $ -1.75 \pm 0.09$ & $  0.50  $ & $-1.55 \pm 0.10$ \\
LG04c\_006479 & $ -1.51 \pm 0.22$ & $ -1.75 \pm 0.09$ & $  0.24  $ & $-1.65 \pm 0.10$ \\
LG04c\_006573 & $ -1.52 \pm 0.10$ & $ -1.42 \pm 0.09$ & $ -0.10  $ & $-1.45 \pm 0.10$ \\
LG04c\_006593 & $ -1.34 \pm 0.$20 & $ -1.36 \pm 0.20$ & $  0.02  $ & $-1.39 \pm 0.11$ \\
LG04c\_006601 & $ -1.52 \pm 0.$31 & $ -1.53 \pm 0.09$ & $  0.01  $ & $-1.54 \pm 0.10$ \\
LG04c\_006788 & $ -1.08 \pm 1.07$ & $ -1.42 \pm 0.13$ & $  0.34  $ & $-1.38 \pm 0.13$ \\
LG04c\_007260 & $ -1.58 \pm 0.09$ & $ -1.49 \pm 0.09$ & $ -0.09  $ & $-1.52 \pm 0.10$ \\
LG04d\_003625 & $ -2.01 \pm 0.20$ & $ -1.60 \pm 0.34$ & $ -0.41  $ & $-1.68 \pm 0.18$ \\
LG04d\_004311 & $ -1.90 \pm 0.53$ & $ -2.14 \pm 0.28$ & $  0.24  $ & $-1.85 \pm 0.12$ \\
LG04d\_004565 & $ -1.65 \pm 0.69$ & $ -2.06 \pm 0.14$ & $  0.41  $ & $-2.05 \pm 0.14$ \\
car1\_t100 & $ -2.12 \pm 0.15$ & $ -1.90 \pm 0.08$ & $ -0.22  $ & $-2.01 \pm 0.11$ \\
car1\_t102 & $ -1.62 \pm 0.10$ & $ -1.53 \pm 0.09$ & $ -0.09  $ & $-1.55 \pm 0.10$ \\
\hline
\end{tabular}
\end{center}
\end{table}

\begin{sidewaystable}
\begin{center}
\caption{Derived properties of member stars in Carina}
\begin{tabular}{cccrccccccr}
\hline
\hline
Star\footnote{\hspace{-4.8cm}\begin{minipage}{20cm}{The nomenclature is such that LG04a--d designates the EIS-fields targeting Carina (EIS-team, priv. comm.) 
followed by the number
 in the respective input catalog.}\end{minipage}} & $\alpha$ (J2000) & $\delta$ (J2000) & r\,[$\arcmin$] & V & \bv & $\Sigma$\,W\,[\AA]  & [Fe/H]$_{R97,ZW}$ & [Fe/H]$_{R97,CG}$  & $\sigma$\,[Fe/H] & $\log p$ \\
\hline
LG04a\_00002 & 06  42  59 & $-$51  00   06 & 15.4 &   19.50 &    0.86 &    3.81  &   $-$1.82      &   $-$1.62  &   0.16 & $-$1.31\\ 
LG04a\_00008 & 06  42  31 & $-$50  59   59 & 10.6 &   18.55 &    1.13 &    4.01  &   $-$1.92      &   $-$1.74  &   0.09 & $-$1.31\\ 
LG04a\_00017 & 06  42  26 & $-$50  59   43 & 09.6 &   19.71 &    0.94 &    2.72  &   $-$2.11      &   $-$1.99  &   0.39 & $-$1.32\\ 
LG04a\_00030 & 06  41  58 & $-$50  59   32 & 04.8 &   17.93 &    1.27 &    4.29  &   $-$1.94      &   $-$1.77  &   0.10 & $-$1.27\\ 
LG04a\_00046 & 06  40  52 & $-$50  59   12 & 07.4 &   19.56 &    0.83 &    2.30  &   $-$2.27      &   $-$2.18  &   0.20 & $-$1.46\\ 
LG04a\_00064 & 06  42  52 & $-$50  58   48 & 13.4 &   19.86 &    0.84 &    4.08  &   $-$1.68      &   $-$1.45  &   0.33 & $-$1.31\\ 
LG04a\_00111 & 06  41  04 & $-$50  57   41 & 05.8 &   19.57 &    0.83 &    2.68  &   $-$2.15      &   $-$2.03  &   0.47 & $-$12.92\\ 
LG04a\_00115 & 06  42  08 & $-$50  57   32 & 05.2 &   19.70 &    0.90 &    2.63  &   $-$2.14      &   $-$2.02  &   0.11 & $-$1.30\\ 
LG04a\_00160 & 06  41  56 & $-$50  56   51 & 03.3 &   19.29 &    0.92 &    3.17  &   $-$2.05      &   $-$1.91  &   0.10 & $-$1.55\\ 
LG04a\_00189 & 06  41  27 & $-$50  56   05 & 03.6 &   18.84 &    1.02 &    4.04  &   $-$1.86      &   $-$1.67  &   0.21 & $-$1.29\\ 
\hline
\end{tabular}
\end{center}
\hspace{2.3cm}\begin{minipage}{20cm}{{\sc note} --- This Table is published in its entirety in the electronic edition of the {\it Astronomical Journal}. 
A portion is shown here for guidance regarding its form and content.
$r$ and $\Sigma$\,W denote the elliptical radius and the CaT linestrength, respectively. 
[Fe/H] and the respective uncertainty
$\sigma$ are given following eqs. 1,2. Finally, $p$ denotes the targets' membership probability.}
\end{minipage}
\end{sidewaystable}

\end{document}